\documentclass[twocolumn, float fix, prb, aps, showpacs, longbibliography]{revtex4-2}
\usepackage{graphicx, amssymb, color, amsmath}
\usepackage{nicefrac}
\usepackage{multirow, array, booktabs}
\usepackage{romannum}
\usepackage{mathtools}
\usepackage{bbm}
\usepackage{mathrsfs}
\usepackage{IEEEtrantools}
\usepackage{caption}
\usepackage{subcaption}
\usepackage[font=small,labelfont=bf,
justification=raggedright,
format=plain]{caption}
\usepackage[titletoc, title]{appendix}
\usepackage[colorlinks, bookmarks=true, citecolor=blue, linkcolor=red, urlcolor=blue]{hyperref}
\usepackage{orcidlink}
\usepackage{braket}
\usepackage{float}
\usepackage{bm}
\usepackage{tikz}
\usepackage{pgfplots}
\usepackage{pgf}
\usepackage{diagbox}
\usepackage{lipsum} 
\AtBeginDocument{\pagenumbering{arabic}}

\begin{document}
\title{Interplay of superconducting, metallic, and crystalline states of composite fermions at $\nu{=}1/6$ in wide quantum wells}
	\author{Ajit C. Balram$^{1,2}$\orcidlink{0000-0002-8087-6015}, Anirban Sharma$^{3}$,  and J. K. Jain$^{3}$\orcidlink{000-0003-0082-5881}}
	\affiliation{$^{1}$Institute of Mathematical Sciences, CIT Campus, Chennai, 600113, India}
	\affiliation{$^{2}$Homi Bhabha National Institute, Training School Complex, Anushaktinagar, Mumbai, 400094, India}
	\affiliation{$^{3}$Department of Physics, 104 Davey Lab, Pennsylvania State University, University Park, Pennsylvania 16802, USA}
	\date{\today}

\begin{abstract} 
		Evidence for developing fractional quantum Hall effect (FQHE) at filling fraction $\nu{=}1/6$ and $1/8$ has recently been reported in wide GaAs quantum wells [Wang \emph{et al.}, Phys. Rev. Lett. {\bf 134}, 046502 (2025)]. In this article, we theoretically investigate the nature of the state at $\nu{=}1/6$ as a function of the quantum well width and the density by considering composite-fermion (CF) crystals, CF Fermi sea, and various kinds of paired CF states. The $f$-wave paired state has the lowest energy among the paired CF states. However, for parameters of interest, the energies of the CF crystal, the CF Fermi liquid, and the $f$-wave paired CF state are too close to call. We, therefore, predict that {\it if} the FQHE at $\nu{=}1/6$ is experimentally confirmed, this state would be an $f$-wave paired state of CFs, which can be verified by measurement of its thermal Hall conductance. Exact diagonalization studies on clean systems with up to 8 electrons show that the ground states at $\nu{=}n/(6n{\pm} 1)$ are incompressible for all widths and densities we have considered and well described by the corresponding Laughlin and Jain states. We propose a phase diagram for large quantum well widths and densities in which at zero disorder, incompressible FQHE states are stabilized at $\nu{=}n/(6n{\pm} 1)$ and $\nu{=}1/6$, but in between these fillings the CF crystal is stabilized. We also present a qualitative discussion on the effects of disorder and propose a schematic phase diagram based on it. With disorder, which creates a spatial variation in the filling factor, two regimes are identified: (i) for small disorder, when the incompressible states percolate at the special fillings, FQHE with quantized Hall plateaus and vanishing longitudinal resistance should occur; and (ii) for larger disorder, when the CF crystal percolates, the longitudinal resistance rises with decreasing temperature but the domains of FQHE liquid produce minima at the special filling factors. The experiments are consistent with the latter scenario. We also mention a possible connection of the phase diagram presented here to a puzzling behavior observed for the fractional quantum anomalous Hall effect in pentalayer graphene.
\end{abstract}
\maketitle
	
\section{Introduction}

The phenomenology of two-dimensional electrons exposed to a strong magnetic field is understood as a consequence of the formation of composite fermions (CFs), which are bound states of electrons and an even number of quantized vortices~\cite{Jain89}. The residual interaction between CFs is weak, and it is a good first approximation to neglect it altogether. In this approximation, the CF theory predicts fractional quantum Hall effect (FQHE) at the odd-denominator fractions $\nu{=}n/(2pn{\pm} 1)$~\cite{Jain89, Jain07, Jain20}, where $n$ and $p$ are positive integers, and CF Fermi seas at even-denominator fractions $\nu{=}1/2p$~\cite{Halperin93, Halperin20, Halperin20b, Shayegan20}. No FQHE occurs at these even-denominator fractions because of the absence of a gap. Indeed, the states at $\nu{=}1/2$ and $\nu{=}1/4$ have been confirmed to be CF Fermi seas~\cite{Willett93, Kang93, Goldman94, Shayegan20}.

However, the CFs do inherit a weak residual interaction from the interaction between electrons, which can sometimes produce qualitatively new physics. In particular, this interaction can sometimes be attractive and cause pairing instability in the CF Fermi sea, which opens a gap leading to even-denominator FQHE. FQHE at $\nu{=}5/2$ was observed in the late 1980s~\cite{Willett87} and is understood as a topological $p$-wave paired state of CFs~\cite{Moore91, Read00}. FQHE at $\nu{=}1/2$ was observed in wide quantum wells (QWs) in the early 1990s~\cite{Suen92, Suen92b}. It was at first interpreted as a bilayer Halperin 331 state~\cite{Halperin83}, but its origin, especially whether it is a one- or a two-component state, remained controversial~\cite{Greiter92, He93, Suen94b, Papic10, Peterson10, Liu14d, Thiebaut15, Mueed15, Mueed16}. Recent theoretical papers have predicted that it is a single-component $p$-wave paired state of CFs~\cite{Zhu16, Sharma23} consistent with very recent experimental observations~\cite{Singh23}. The physical picture here is that the width of the quantum well suppresses the short-range part of the electron-electron interaction, resulting in the pairing of CFs. A FQHE at $\nu{=}1/4$ was also observed in wide QWs~\cite{Shabani09a, Shabani09b, Drichko19} and has been predicted to be an $f$-wave paired state of CFs~\cite{Faugno19, Sharma23}. Many other even-denominator FQHE states have been observed in recent years. In hole-based systems, even-denominator FQHE states have been observed at $\nu{=}1/4$, $\nu{=}3/4$, and $\nu{=}1/6$ in narrow QWs~\cite{Wang23, Wang22a, Wang25a}. These have been attributed to Landau level (LL) mixing, which is large in hole-type systems due to the large effective mass of the holes. A theoretical study incorporating LL mixing in a fixed-phase diffusion Monte Carlo method supports this picture~\cite{Zhao23}. Wang {\it et al.}~\cite{Wang23b} have seen FQHE also at $\nu{=}3/8$ and $\nu{=}3/10$ in hole type systems; in the absence of LL mixing, the states at $3/8$ and $3/10$ were predicted to arise from an anti-Pfaffian state of CFs~\cite{Mukherjee12, Mukherjee14c, Mukherjee15b}. FQHE has been seen also at $\nu{=}2{+}3/8$~\cite{Pan08, Kumar10, Bellani10} and been studied theoretically~\cite{Hutasoit16, Balram21}. FQHE at half-filling has also been observed in the $\mathcal{N}{=}3$ LL of monolayer graphene~\cite{Kim19}. Theoretical calculations~\cite{Kim19, Sharma22} suggest that the leading candidate to describe this FQHE state is the Jain-$221$ parton state~\cite{Jain89b}, which represents an $f$-wave paired state of CFs~\cite{Wen91, Balram18}. FQHE has been observed in the half-filled $\mathcal{N}{=}1$ LLs of bilayer graphene~\cite{Zibrov16, Li17, Huang21, Assouline23, Kumar24} and trilayer graphene~\cite{Chen24}; these are believed to be analogous to the $5/2$ FQHE~\cite{Zhu20a, Balram21b}. Even-denominator FQH phases at $\nu{=}3/4$ in the $\mathcal{N}{=}0$ LL of bilayer graphene~\cite{Kumar24} and at $\nu{=}1/2$ in the $\mathcal{N}{=}0$ LL of trilayer graphene~\cite{Chanda25} were recently observed. These likely involve LL mixing, but a satisfactory understanding of the underlying physical mechanism for these states' incompressibility is lacking~\cite{Yutushui25, Huang25}. FQHE states have been observed at an isospin transition in the $\mathcal{N}{=}0$ LLs of monolayer graphene at half- and quarter-filling and their origin was attributed to two-component physics~\cite{Zibrov17, Wu22}. 
%%%%%%%%%%%%%%%%%%%%%%%%%%%%%%%%%%%%%%%%%%%%%%%%%%%%%%%%%%%%%%%%%%%%%%%%%%%%%
	\begin{figure}[t]
		\begin{center}
			\includegraphics[width=0.49\textwidth,height=0.32\textwidth]{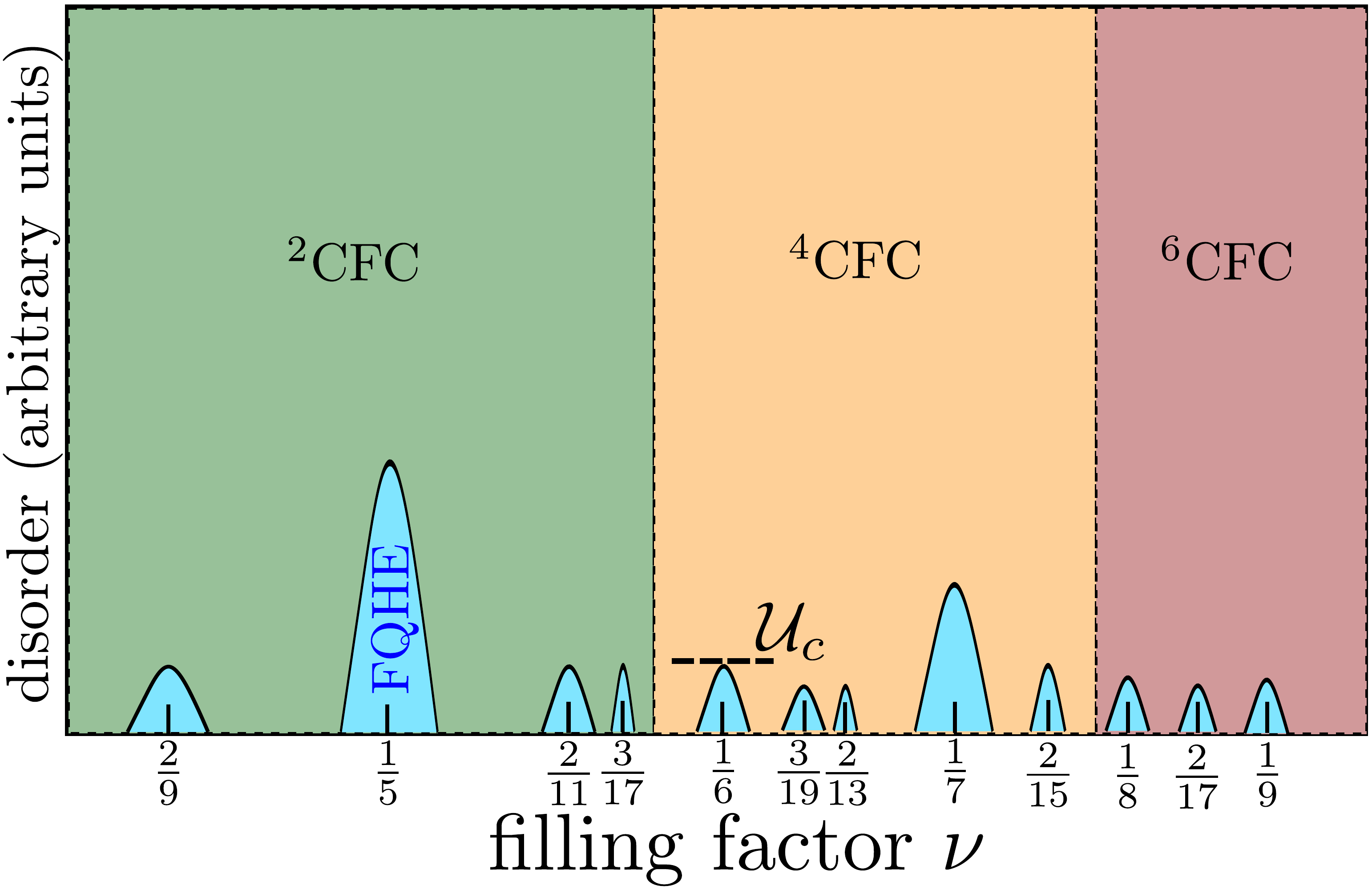}	\\
                \includegraphics[width=0.49\textwidth,height=0.17\textwidth]{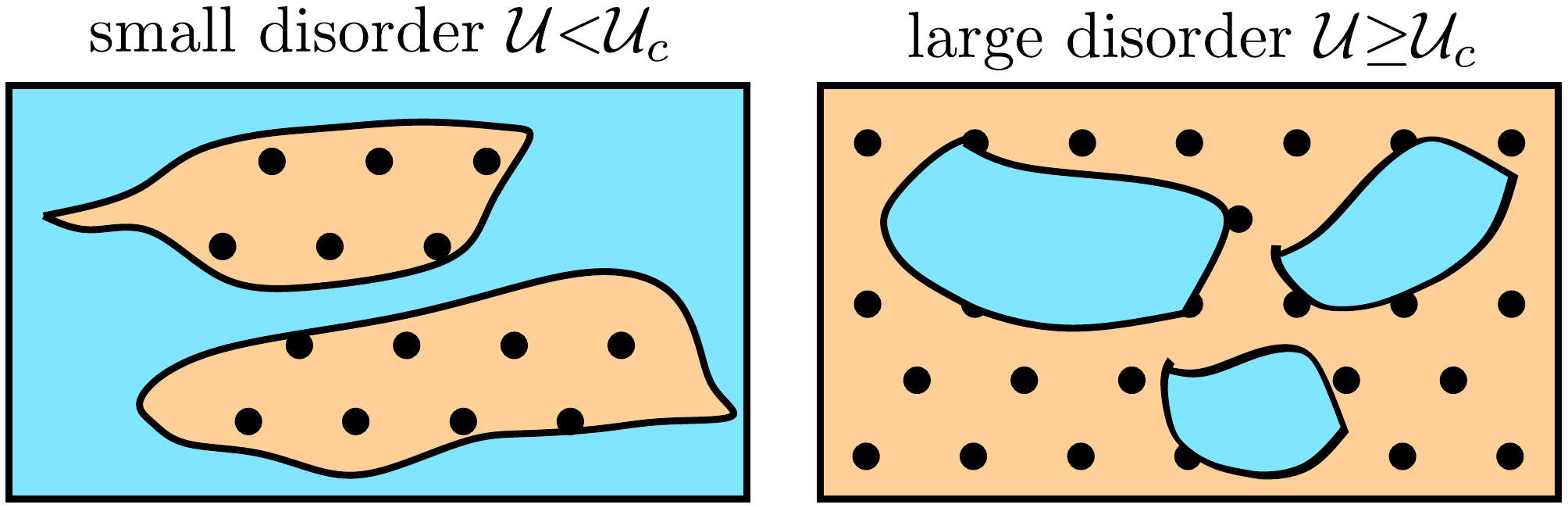}	\\
			\caption{(color online) Top panel shows the proposed schematic phase diagram as a function of filling factor and disorder at large quantum-well widths and densities. This is identical to that shown in Ref.~\cite{Zuo20} except that there are incompressible states also at $\nu{=}1/6$ and $\nu{=}1/8$. The phase labeled CFC (CF crystal) is a perfect crystal only at zero disorder; the correlation length of crystalline order decreases with increasing disorder. The bottom panels show two possible phases at a given filling factor (such as $n/(6n{\pm}1)$ or $1/6$) as a function of disorder. The left panel shows the ``FQHE phase" which appears for low disorder (${\cal U}{<}{\cal U}_c$) when the incompressible liquid (blue) percolates through the sample. The bottom right panel shows the ``correlated mixed-phase insulator" which appears for large disorder (${\cal U}{\geq}{\cal U}_c$). Here, the crystal (yellow) percolates but contains puddles of the FQHE liquid, which diminishes the longitudinal resistance. The critical disorder ${\cal U}_c$ for a given fraction corresponds to the height of the dome in the phase diagram (marked for $\nu{=}1/6$).}
			\label{fig: schematic_crystal_liquid_disorder}
		\end{center}
	\end{figure}
	%%%%%%%%%%%%%%%%%%%%%%%%%%%%%%%%%%%%%%%%%%%%%%%%%%%%%%%%%%%%%%%%%%%%%%%%%%%%%

The motivation for the present study comes from a recent experiment that has reported evidence for developing FQHE at $\nu{=}1/6$ and $\nu{=}1/8$ in wide QWs, in addition to many Jain-sequence states at $\nu{=}n/(6n{\pm} 1)$ and $\nu{=}n/(8n{\pm} 1)$~\cite{Wang25}. These manifest through deep minima in $R_{xx}$ riding on top of a background resistance that rapidly rises with decreasing temperature. Evidence for the $n/(6n{\pm} 1)$ states riding on a large $R_{xx}$ background was reported more than two decades ago~\cite{Pan02}. This was surprising because early theoretical studies had suggested~\cite{Lam84} that for $\nu{<}1/6.5$ a Wigner crystal (WC) is stabilized rather than the FQHE liquid. Subsequent theoretical studies found that many types of CF crystals (CFCs) can occur and demonstrated a rather intricate interplay between the liquid and CF crystal states~\cite{Chang05, Archer13, Zhao18}. A theoretical study argued that the FQHE states survive down to much lower filling factors, although they are separated by CF crystals in between~\cite{Zuo20}. 

This article considers many candidate states at $\nu{=}1/6$ in a wide QW, including the CF Fermi sea, various kinds of crystals, and several kinds of paired CF states. The finite width alters the electron-electron interaction, which we obtain by evaluating the transverse wave function in a local density approximation (LDA)~\cite{Ortalano97, Martin20}. We do not consider LL mixing, which is not expected to be relevant here since the LL mixing parameter is small in this electron-based system. Our primary conclusions are as follows:

\subsection{Summary of primary conclusions}

Among the various FQHE states we have considered at $\nu{=}1/6$, the Jain-2211111 (or 2$^2$1$^5$) parton state~\cite{Jain89b}, defined below, is the most favorable. This state represents an $f$ wave pairing of CFs~\cite{Balram18} and is predicted to support excitations with non-Abelian statistics~\cite{Wen91}. However, for the QW widths and densities corresponding to the experiment~\cite{Wang25}, the energies of the CF Fermi sea, the $^4$CF crystal ($^4$CFC) [i.e., a crystal of CFs with 4 vortices bound to them], and the $f$-wave paired CF states are too close to call. 
  
Based on these calculations, we cannot conclude that the $f$-wave paired CF state will occur for large QW widths and densities. However, {\it if} the FQHE at $\nu{=}1/6$ is experimentally confirmed in these systems, we predict that it will be an $f$-wave paired state of CFs. This can be verified experimentally, for example, by determining the thermal Hall conductance of this state. The thermal Hall conductance is given by $\kappa_{xy}{=}c_{-}\pi^2k_B^2T/3h$, where $c_{-}$ is the chiral central charge~\cite{Kane97}; for state with CF pairing in relative angular momentum $l$ channel, we have $c_{-}{=}1{+}l/2$ and in particular, for $f$ wave pairing ($l{=}3$) of CFs, we have $c_{-}{=}5/2$~\cite{Wen91, Balram19, Faugno19}.

We have also explored FQHE states away from $\nu{=}1/6$. Exact diagonalization (ED) calculations on systems with up to 8 electrons show that the $1/7$ Laughlin and the $2/11$ and $2/13$ Jain states remain stable for the parameters of interest. 

\subsection{Proposed Phase Diagram}

Combining with experiments, we propose the schematic phase diagram depicted in the upper panel of Fig.~\ref{fig: schematic_crystal_liquid_disorder} for large QW widths, which is a slight modification of that presented by Zuo {\it et al.}~\cite{Zuo20}. At zero disorder, the Jain states are stabilized at and near $\nu{=}n/(6n{\pm} 1)$, and the $f$-wave paired CF state is stabilized at and near $\nu{=}1/6$. At filling factors in between, the CF crystal is stabilized. Now consider disorder, which causes spatial variations in the filling factor. Due to the intervening crystal phases, the behavior is very different than that observed for, say, the $\nu=n/(2n+1)$ FQHE states. Specifically, in the vicinity of $\nu{=}n/(6n{\pm} 1)$ or $\nu{=}1/6$, we can expect two distinct phases: 

(i) {\bf FQHE phase}: At weak disorder, some domains of CF crystal appear, but the incompressible FQHE state continues to percolate (lower left panel of Fig.~\ref{fig: schematic_crystal_liquid_disorder}). This phase would result in an FQHE plateau with a vanishing longitudinal resistance. 

(ii) {\bf Correlated mixed-phase insulator}: Above some critical disorder, the CF crystal begins to percolate but contains puddles of the incompressible liquid (lower right panel of Fig.~\ref{fig: schematic_crystal_liquid_disorder}). This system is insulating due to the pinning of the crystal, but, unlike for an uncorrelated insulator, the presence of FQHE domains will reduce the resistance at and around these special filling factors relative to the crystal state formed away from the special fillings where the crystal does not contain any FQHE domains, thus producing minima in the resistance at the special fillings. (These statements apply to the resistance at a non-zero temperature; the resistance is infinite at zero temperature.)

The observed behavior in Ref.~\cite{Wang25} is consistent with the correlated mixed-phase insulator, and we are not aware of any other plausible explanation for it. This phase diagram predicts that with a sufficient reduction in disorder, ideal FQHE would be seen at $\nu{=}n/(6n{\pm} 1)$ and $\nu{=}1/6$ with insulating states in between. We note that similar physics has been seen at $\nu{=}1/5$, which at first appeared as a minimum on a rising background and turned into an FQHE state only with significant improvement in the sample quality~\cite{Pan99}; the insulating state, presumably a pinned CF crystal, has persisted between $\nu{=}1/5$ and $\nu{=}2/9$ and also below $\nu{=}1/5$. 
We expect analogous physics for $\nu{=}1/8$ and the FQHE states surrounding it. However, we have not studied this filling factor regime quantitatively.

\subsection{Possible relation to other work}

\subsubsection{Superconductor-metal-insulator transition}
The transition from a two-dimensional superconductor to an insulator with increasing disorder or the application of a magnetic field has been of great interest and served as a canonical example of a quantum phase transition~\cite{Fisher89, Fisher89a, Haviland89, Hebard90, Dubi07, Goldman10}. It has been thought that a metallic state may exist at the critical point~\cite{Imada98, Kapitulnik19}. Analogous questions may be asked about a ``CF superconductor" to a CF insulator transition as a function of disorder (for an early study see Refs.~\cite{Kalmeyer92, Kalmeyer93}). In this article we perform a variational study (in the absence of disorder) and find the possibility of a transition, as a function of the QW width, from a CF crystal (which would manifest as an insulator in the presence of disorder) into a ``CF superconductor"; this transition may or may not go through the metallic CF Fermi sea phase. 

\subsubsection{Fractional quantum anomalous Hall effect in pentalayer graphene}
Recently, fractional quantum anomalous Hall effect (FQAHE), i.e., FQHE in zero magnetic field, has been reported by several groups~\cite{Cai23, Fan23, Park23, Zeng23, Lu24}. This effect occurs due to the presence of flat bands with non-zero Chern numbers that mimic a Landau level. The band filling plays the role of LL filling. Some of the ideas presented in this work may be relevant to the puzzling behavior observed in pentalayer graphene~\cite{Lu24, Lu25}. In Ref.~\cite{Lu24}, they see FQAHE at several Jain fractions (2/3, 3/5, 4/7, 4/9, 3/7, and 2/5). In a subsequent paper~\cite{Lu25}, the authors find that going to yet lower temperatures, the FQAHE state at 2/3 disappears, leaving behind only the integer quantum anomalous Hall effect there (the other FQAHE states remain robust to the lowest temperatures). We suggest that one possible way to rationalize the observation is to imagine that the 2/3 FQAHE state is separated by crystals, as shown in the phase diagram of Fig.~\ref{fig: schematic_crystal_liquid_disorder}. Furthermore, we expect that the disorder effectively becomes weaker with increasing temperature, as carriers begin to move and screen the disorder potential~\cite{DasSarma24a}. Alternatively, the crystal may melt at some temperature. Thus, an insulator at very low temperatures can become an FQAHE state at a somewhat elevated temperature, as seen in experiments. (Of course, eventually, the FQAHE will also be destroyed for sufficiently high temperatures.) We stress that this is only a scenario, and much further work will be required for its verification. We also note that, very recently, signatures of superconductivity have also been observed in this pentalayer graphene system~\cite{Han25}.

The plan for the rest of the paper is as follows. Section~\ref{sec: background} contains some theoretical background, followed by Sec.~\ref{sec: wf_candidates}, which lists all candidate states that we have studied. We provide results from ED and variational calculations in the two subsequent sections, Secs.~\ref{sec: ED} and~\ref{sec: VMC}, respectively, and discuss their implications in Sec.~\ref{sec: discussion}.

\section{Background}
\label{sec: background}

In this section, we mention certain standard concepts and states that are useful later.

\underline{CF theory}: The wave functions for the ground states at $\nu{=}n/(2pn{\pm}1)$ are given by $\Psi_{n/(2pn{\pm}1)}^{\rm Jain} {=} \mathcal{P}_{\rm LLL}\Phi^{2p}_{1}\Phi_{{\pm}n}$, where $\Phi_n$ and $\Phi_{{-}n}{=}[\Phi_{n}]^{*}$ are the Slater determinant wave functions of the states with $n$ filled LLs in positive and negative magnetic fields. Here $\mathcal{P}_{\rm LLL}$ projects the state into the lowest LL (LLL). For the ground state at $\nu{=}1/(2p{+}1)$ the  wave function reduces to the Laughlin wave function~\cite{Laughlin83}. 

\underline{Parton theory}: Also relevant below is the parton theory of the FQHE~\cite{Jain89b}. In the simplest version of this approach, one constructs candidate FQHE states by dividing electrons into an odd number of fictitious particles called partons, placing each parton species into an integer quantum Hall state $\Phi_{n_{\lambda}}$, and finally combining the partons back into physical electrons. The resulting state, denoted as the Jain-$n_1n_2{\cdots}n_k$ parton state, is given by
\begin{equation}
\Psi^{n_1n_2....n_k}=\mathcal{P}_{\rm LLL} \prod_{\lambda = 1}^{k}\Phi_{n_{\lambda}}(\{z_i\}),
\end{equation}
where $z_{i}$ is the two-dimensional coordinate of the $i^{\rm th}$ electron parameterized as a complex number. Requiring that each parton species occupy the same area imposes the condition that the charge of the $\lambda$ parton $e_{\lambda}{=}\nu/n_{\lambda}$ (in units of electron charge) with $\sum_{\lambda} e_{\lambda}{=}1$. The filling factor of the state is thus given by $\nu{=}[\sum_{\lambda{=}1}^k n_{\lambda}^{{-}1}]^{{-}1}$. Note that $n$ can take a negative integer value.

\underline{Spherical geometry}: All our computations are carried out on the Haldane sphere~\cite{Haldane83} with $N$ electrons moving on its surface. A magnetic monopole of strength $2Q$ placed at the sphere's center produces an outward radial magnetic flux of strength $2Qhc/e$. Appropriate to the spherical geometry, the total orbital angular momentum $L$ and its $z$-component $L_{z}$ are good quantum numbers. Incompressible quantum Hall states occur when the flux $2Q$ is related to $N$ as $2Q{=}\nu^{{-}1}N{-}\mathcal{S}$, where $\mathcal{S}$, called the shift, is a topological quantum number that characterizes certain features of the FQHE state~\cite{Wen92}. For the parton states given above, the shift is given by $\mathcal{S}{=} \sum_{\lambda{=}1}^{k}n_{\lambda}$. Quantum Hall states and the CFFS are uniform on the sphere and thus have $L{=}0$ while, generically, the crystal has $L{\neq}0$. The numerical computations are carried out in disorder-free clean systems, and the effects of LL mixing are neglected. 

The electronic coordinates on the sphere can be parametrized in terms of spinor coordinates~\cite{Haldane83} $u_{k}{=}\cos(\theta_{k}/2)e^{i\phi_{k}/2}$ and $v_{k}{=}\sin(\theta_{k}/2)e^{-i\phi_{k}/2}$, where $\theta_{k}$ and $\phi_{k}$ are the polar and azimuthal angles of the $k^{\rm th}$ electron on the sphere. In terms of the spinor coordinates, the wave function of the filled LLL is given by $\Phi_{1}{=}\prod_{1{\leq} i{<}j{\leq}N}( u_{i} v_{j} {-} u_{j} v_{i})$.

	\underline{Finite width}: Finite QW width essentially changes the interaction between electrons to $\int dw_1 \int dw_2 \frac{|\zeta(w_1)|^{2}|\zeta(w_2)|^{2}}{\sqrt{r^2{+}(w_1{-}w_2)^2}}$, where $r$ is the distance between the electrons within the plane, $w_1$ and $w_2$ are the coordinates in the transverse direction and $\zeta(w)$ is the transverse wave function. We obtain the transverse wave function $\zeta(w)$ in a local density approximation whereby it is a function of both the well-width and the density~\cite{Martin20}. 
    
    \section{Candidate states at $\nu{=}1/6$}
	\label{sec: wf_candidates}
    
At $\nu{=}1/6$, we consider the following candidate states:
    
\underline{CF Fermi sea (CFFS)}: The wave function for this state is given by:
		\begin{equation}
			\Psi^{\rm CFFS}_{1/6}=\mathcal{P}_{\rm LLL}\Psi^{\rm FS}\Phi^{6}_{1}.
			\label{eq: wf_CFFS}
		\end{equation}
		This state occurs at $2Q{=}6N{-}6$. On the sphere, we shall consider filled-shell CF states that are known to provide a good representation of the CFFS~\cite{Rezayi94, Balram17, Liu20}. Variants of the CFFS wave function can also be constructed by projecting the state into the LLL in different ways to obtain wave functions such as $ \Psi^{\rm CFFS}_{1/6} {=}  \Psi^{\rm CFFS}_{1/4} {\times}  \Phi_1^2$, $ \Psi^{\rm CFFS}_{1/6} {=} \Psi^{\rm CFFS}_{1/2}  {\times}  \Phi_1^4$, etc. We will use the wave function given in Eq.~\eqref{eq: wf_CFFS}, which has been shown to give the minimal Coulomb energy~\cite{Balram15a}.
         
   \underline{$^{2p}$CF crystal}  The wave function of the uncorrelated Hartree-Fock crystal of electrons is given by  $\Psi^{\rm HF-crystal}{=}{\rm Det}[\eta_{\vec{R}_i}(\vec{r}_j)]$ where $\eta_{\vec{R}_i}(\vec{r})$ is a coherent state wave packet localized at $\vec{R}_i$, where $\{\vec{R}_i\}$ form a hexagonal lattice. Because such a lattice cannot be wrapped neatly, i.e., without defects, onto the surface of a sphere, we work with a Thomson lattice~\cite{Thomson04}, which minimizes the energy of $N$ classical particles on the surface of a sphere~\cite{Archer13}. Past calculations have shown that this is a good approximation to the hexagonal lattice in the thermodynamic limit. We consider below a class of $^{2p}$CF crystals given by 
   \begin{equation}
\Psi_Q^{^{2p}\rm CF-crystal}   = \Psi_{Q^*}^{\rm HF-crystal} \Phi_1^{2p}
   \end{equation}
where $2Q{=}2Q^*{+}2p(N{-}1)$. Here, $p$ is treated as a variational parameter. The HF crystal corresponds to $2p{=}0$. For the crystal at $\nu{=}1/6$, we have found that the optimal crystal has $2p{=}4$, referred to as the $^4$CF crystal.

        \underline{Moore-Read Pfaffian}: The wave function of this state is given by~\cite{Moore91}:
		\begin{equation}
			\Psi^{\rm MR-Pf}_{1/6}={\rm Pf}[\left( u_iv_j-v_iu_j  \right)^{-1}] \Phi^{6}_{1},
			\label{eq: wf_Pf}
		\end{equation} 
        The $\nu{=}1/6$ Pfaffian state occurs at $2Q{=}6N{-}7$.

\underline{$2211111$ ($2^{2}1^{5}$) parton state}: The wave function of this state is given by:
		\begin{equation}
			\Psi^{2^{2}1^{5}}_{1/6}=\mathcal{P}_{\rm LLL}\Phi^{2}_{2}\Phi^{5}_{1} \sim [\Psi^{\rm Jain}_{2/5}]^{2}\Phi_{1}.
			\label{eq: wf_2211111}
		\end{equation}
		 In the spherical geometry, the $2^{2}1^{5}$ state occurs at flux $2Q{=}6N{-}9$. The wave functions on the two sides of the $\sim$ sign in Eq.~\eqref{eq: wf_2211111} differ in the details of how the projection into the LLL is carried out. We anticipate that such details only make minor quantitative changes in the wave functions and do not alter the universality class of the underlying state~\cite{Balram15a,  Anand22}. The form of the wave function given on the rightmost side of Eq.~\eqref{eq: wf_2211111} is amenable to numerical calculations via the Jain-Kamilla (JK) projection method~\cite{Jain97, Jain97b, Moller05, Davenport12, Balram15a}. To extract the topological properties of the state given in Eq.~\eqref{eq: wf_2211111}, we note that the unprojected parton states lend themselves to a description in terms of a Chern-Simons field theory~\cite{Wen91}. Moreover, recent numerical calculations~\cite{Anand22} suggest that LLL projection does not alter the topological properties of the state.  Therefore, assuming this to be the case, we can identify all the topological quantum numbers of the parton state using its Chern-Simons field-theoretic description.
		
\underline{$\bar{2}^{2}1^{7}$ parton state}: The wave function of this state is given by:
		\begin{equation}
			\Psi^{\bar{2}^{2}1^{7}}_{1/6}=\mathcal{P}_{\rm LLL}[\Phi^{2}_{2}]^{*}\Phi^{7}_{1} \sim [\Psi^{\rm Jain}_{2/3}]^{2}\Phi^{3}_{1}.
			\label{eq: wf_bar2bar21111111}
		\end{equation}
		This state occurs at $2Q{=}6N{-}3$. This wave function lies in the same universality class as the anti-Pfaffian wave function~\cite{Levin07, Lee07} (the hole partner of the Pfaffian wave function) at $\nu{=}1/6$~\cite{Balram18}.

    Many of the experimentally accessible properties of the above states are tabulated in Table~\ref{tab:one_sixth}.
	
	\begin{table}[h]
		\begin{center}
			\begin{tabular} { | c | c | c | c |}
				\hline
				\multicolumn{4}{|c|}{States at $\nu{=}1/6$}\\
				\hline
				State & wave function & $\mathcal{S}$ & $c_{-}$\\
				\hline
				CF Fermi sea & $\mathcal{P}_{\rm LLL}\Psi^{\rm FS}\Phi_1^6$ & 6 & --\\
				\hline
				Moore-Read Pfaffian & $\text{Pf}\left([u_i v_j - u_j v_i]^{-1}\right)\Phi_1^6$ & 7 & 3/2 \\
				\hline
				$2^{2}1^{5}$ & ${\cal P}_{\rm LLL}\Phi^{2}_{2}\Phi_1^5\sim [\Psi^{\rm Jain}_{2/5}]^{2}\Phi_1$ & 9 & 5/2\\
				\hline
				$\bar{2}^{2}1^{7}$ & ${\cal P}_{\rm LLL}[\Phi^{2}_2]^{*}\Phi_1^7\sim [\Psi^{\rm Jain}_{2/3}]^{2}\Phi^{3}_1$ & 3 & -1/2\\
				\hline
				PH-Pfaffian & $\mathcal{P}_{\rm LLL}\text{Pf}\left([u^{*}_i v^{*}_j-u^{*}_j v^{*}_i]^{-1}\right)\Phi_1^6$ & 5 & 1/2\\
                    \hline
                    (anti-$2^{2}1)1^{4}$ & $\left[{\cal P}_{\rm ph}\left(\Phi^{2}_{2}\Phi_1\right)\right]\times \Phi_1^4$ & 1 & -3/2\\
				\hline
				
			\end{tabular}
		\end{center}
		\caption{\label{tab:one_sixth} Candidate states at $\nu{=}1/6$. Their wave function, shift $\mathcal{S}$ on the sphere, and chiral central charge $c_{{-}}$ are given. The thermal Hall conductance $\kappa_{xy}$ is related to the chiral central charge as $\kappa_{xy}{=}c_{{-}}[\pi^2 k_{\rm B}^2 /(3h)]T$. For the gapless composite fermion Fermi sea, the thermal Hall conductance is not quantized. For completeness, in the last two rows, we also include the so-called particle-hole symmetric Pfaffian (PH-Pfaffian)~\cite{Jolicoeur07, Zucker16} and the (anti-$2^{2}1)1^{4}$ states (anti represents the hole partner state) although we have not considered these in the present work.}
	\end{table}
	
	\section{Exact diagonalization studies}
    \label{sec: ED}

	In this section, we calculate the overlaps and gaps for $N{=}8$ electrons, which is the largest system for which we can construct the wave functions in the Fock space, of the incompressible states discussed in Sec.~\ref{sec: wf_candidates}. To evaluate the Fock-space decomposition of the candidate wave functions, we follow the method in Refs.~\cite{Sreejith13, Balram21}. In this method, we first evaluate all the $L{=}0$ states, \{$\Phi_{L{=}0}^\alpha$\}, of the LLL system. The uniform LLL projected candidate wave function, $\Psi$, can be expressed as $\Psi(\{z_j\}){=}\sum_\alpha C_\alpha \Phi_{L{=}0}^\alpha(\{z_j\})$. By evaluating this equation at different particle configurations $\{z_j\}$ we get a set of linear equations which can be solved to obtain $C_\alpha$. The bottleneck in this procedure is the accuracy of the $L{=}0$ states of the relevant Hilbert space, which in practice limits the system sizes that we can access at $\nu{=}1/6$ to $N{=}8$ electrons.
	
	Aside from the ground state overlaps, we have also computed the gaps to neutral and charged excitations. The neutral gap is defined as the difference in energy between the two lowest-lying states at the value $2Q$ corresponding to the ground state. The charge gap (which can be accessed via activated transport measurements) is defined as the energy needed to create a far-separated pair of fundamental (smallest magnitude charge) quasihole and quasiparticle. For a system of $N$ particles wherein the ground state occurs at a value of flux $2Q$, the charge gap can be obtained as~\cite{Balram20b}:
	\begin{eqnarray}
		\label{eq: charge_gap}
		\Delta^{\rm charge}&=&\frac{\mathcal{E}(2Q-1)+\mathcal{E}(2Q+1)-2\mathcal{E}(2Q)}{n_{q}}, \\
		\mathcal{E}(2Q   )&=&E(2Q)-N^{2} \frac{\mathcal{C}(2Q)}{2}.  \nonumber 
	\end{eqnarray}
	Here $E(2Q)$ is the ground state energy of $N$ particles at flux $2Q$ that is evaluated using ED, $\mathcal{C}(2Q)$ is the average charging energy at flux $2Q$ which accounts for the contribution of the positively charged background, and $n_{q}$ is the number of fundamental quasiparticles (quasiholes) produced upon the removal (addition) of a single flux quantum in the ground state. For the zero-width $1/r$ Coulomb interaction, $\mathcal{C}(2Q){=}1/\sqrt{Q}~e^{2}/(\epsilon\ell)$, where $\epsilon$ is the dielectric constant of the background host material and $\ell{=}\sqrt{\hbar c/(eB)}$ is the magnetic length at magnetic field $B$, and for the finite-width interaction the charging energy has to be numerically evaluated from the pseudopotentials~\cite{Balram20b, Sharma23}. For all the incompressible states at $\nu{=}1/6$ discussed in Sec.~\ref{sec: wf_candidates} we have $n_{q}{=}2$ while for the Jain states at $\nu{=}n/(2pn{\pm}1)$, we have $n_{q}{=}n$.

\begin{figure*}[htpb]
		\begin{center}
			\includegraphics[width=0.32\textwidth,height=0.17\textwidth]{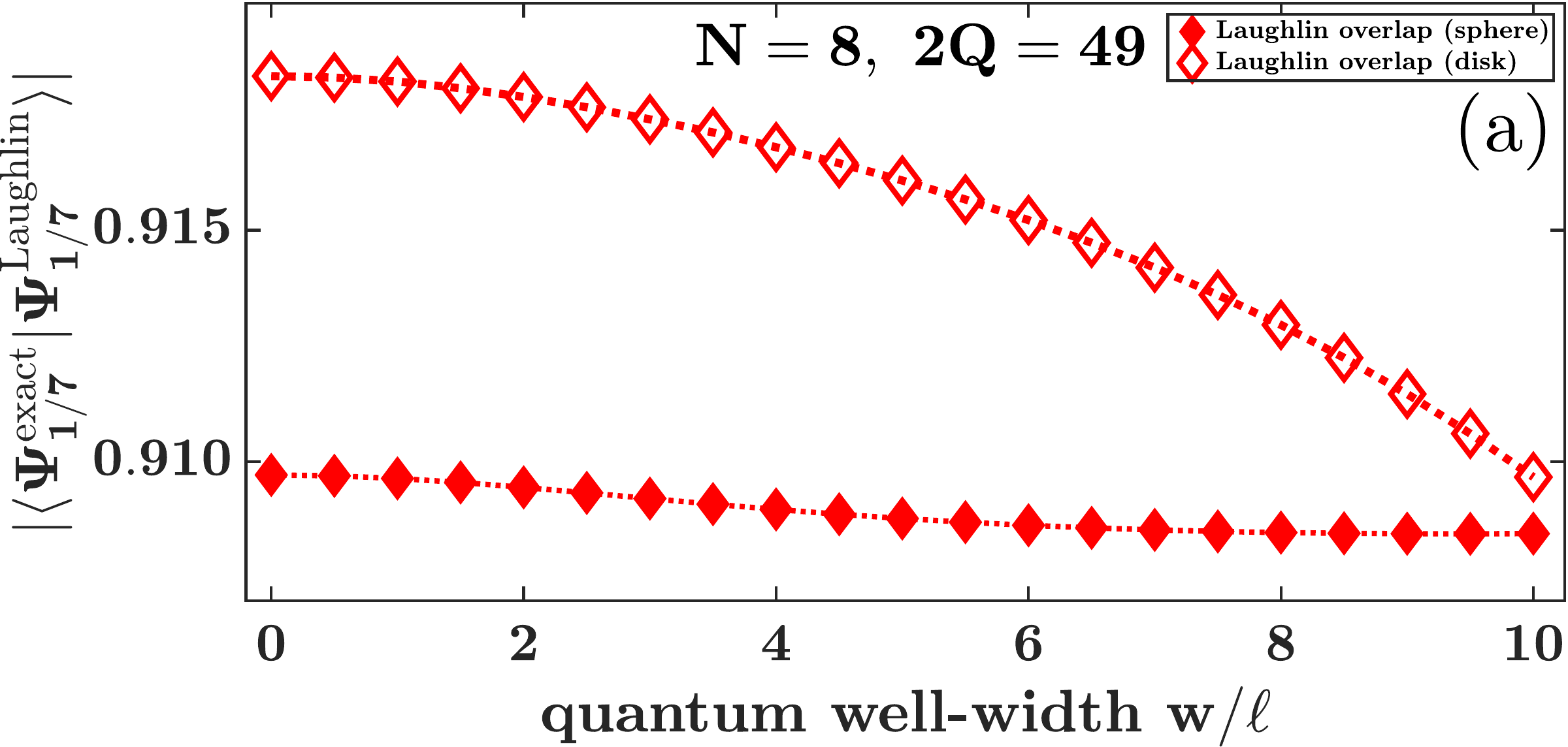}		
			\includegraphics[width=0.32\textwidth,height=0.17\textwidth]{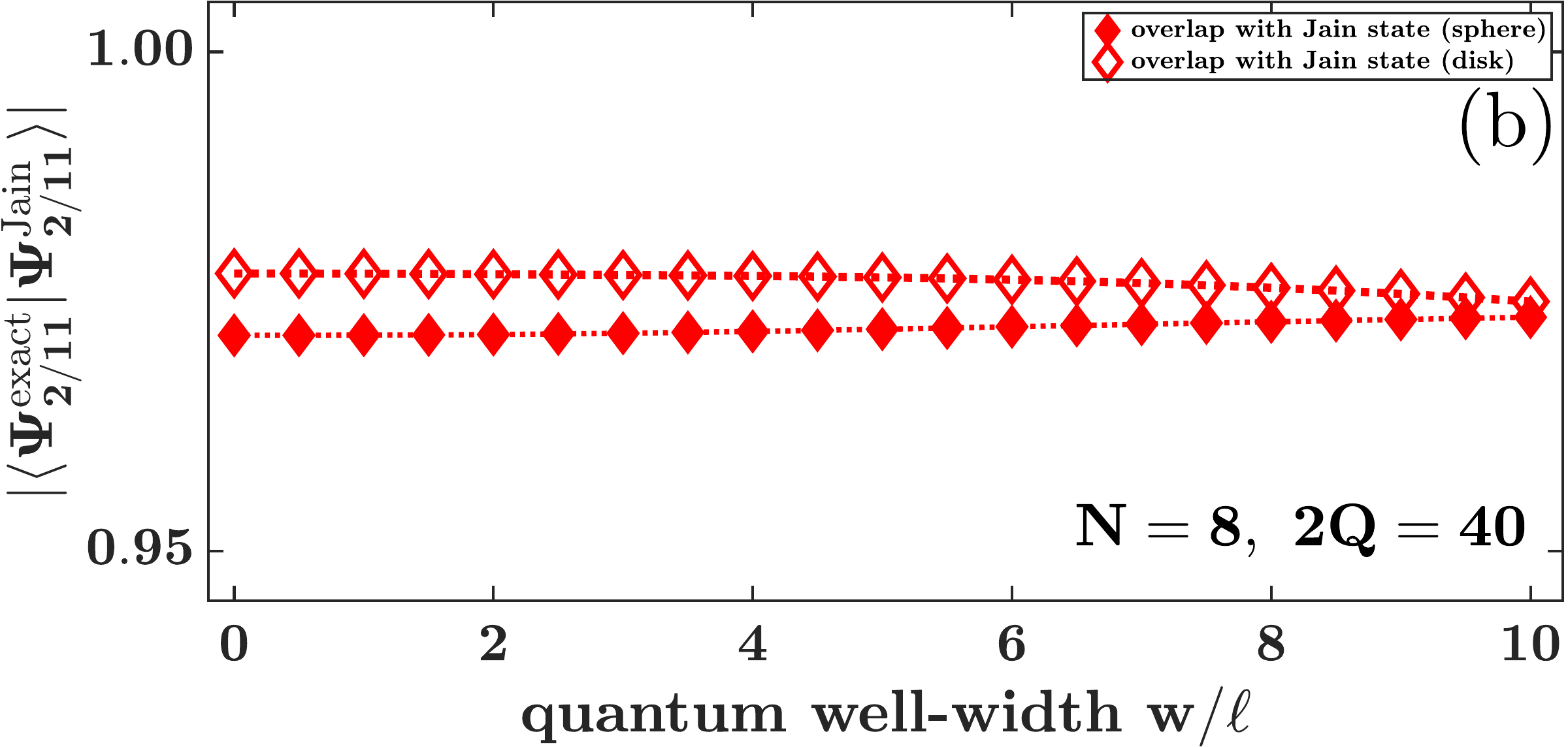} 
			\includegraphics[width=0.32\textwidth,height=0.17\textwidth]{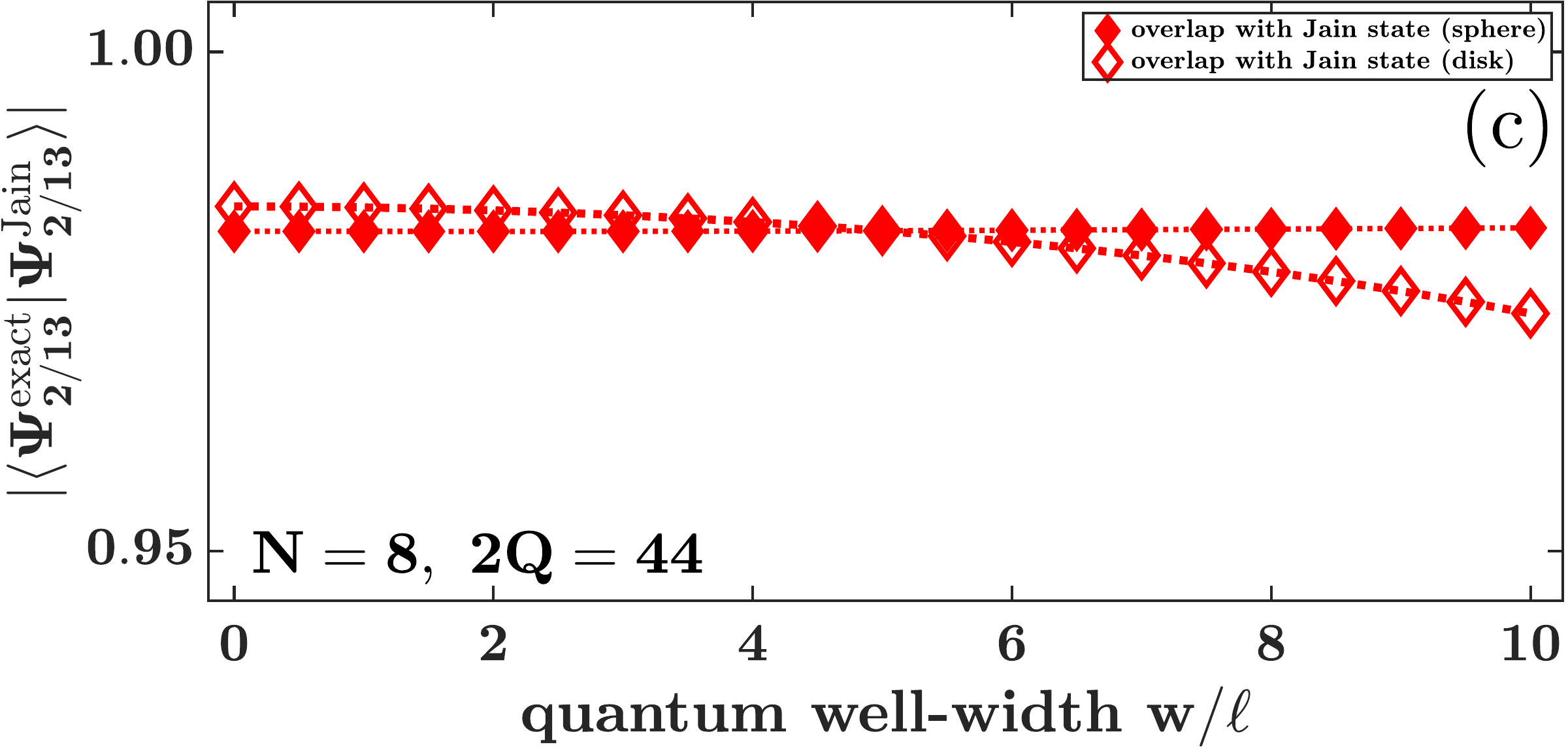} 
			\\
			\includegraphics[width=0.32\textwidth,height=0.17\textwidth]{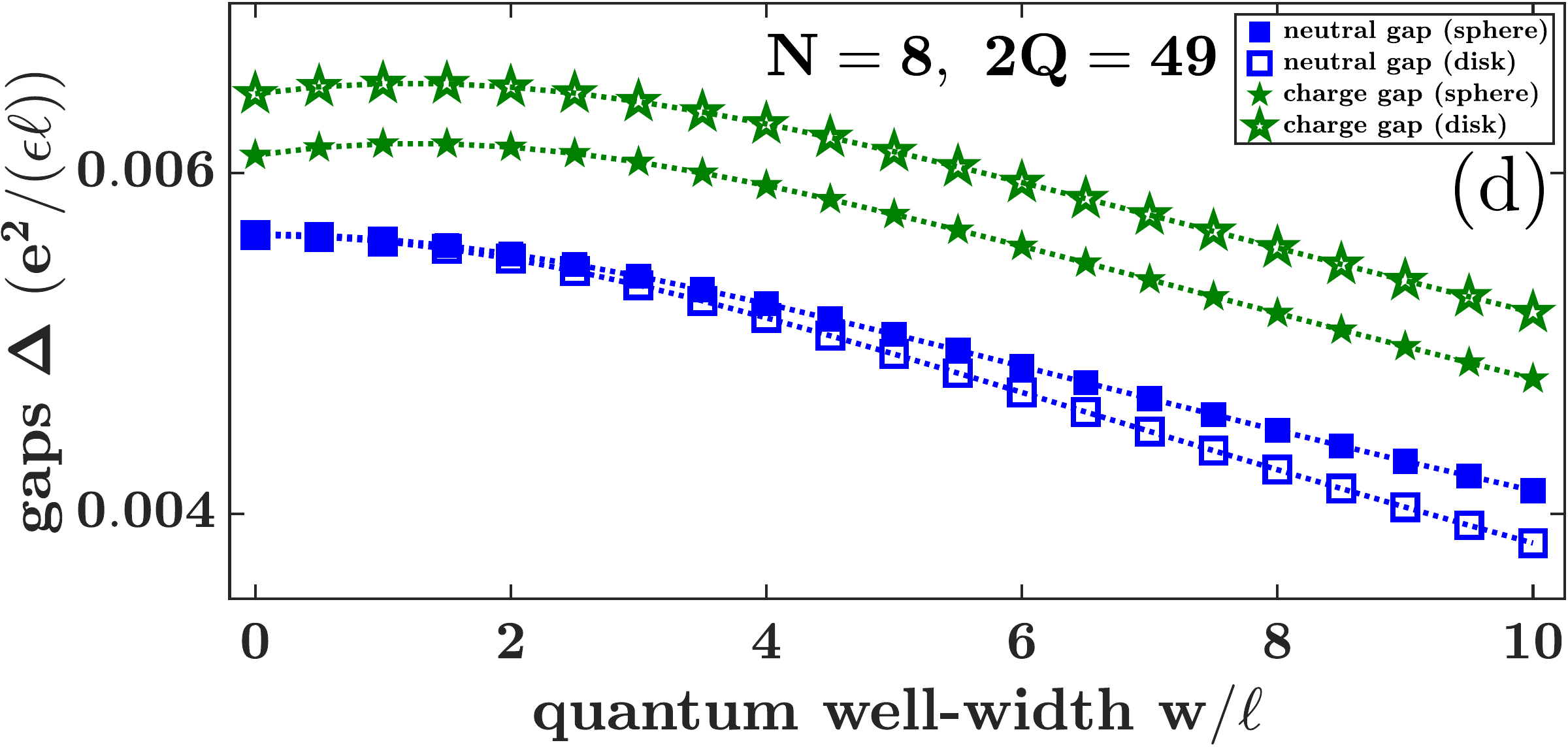} 	
			\includegraphics[width=0.32\textwidth,height=0.17\textwidth]{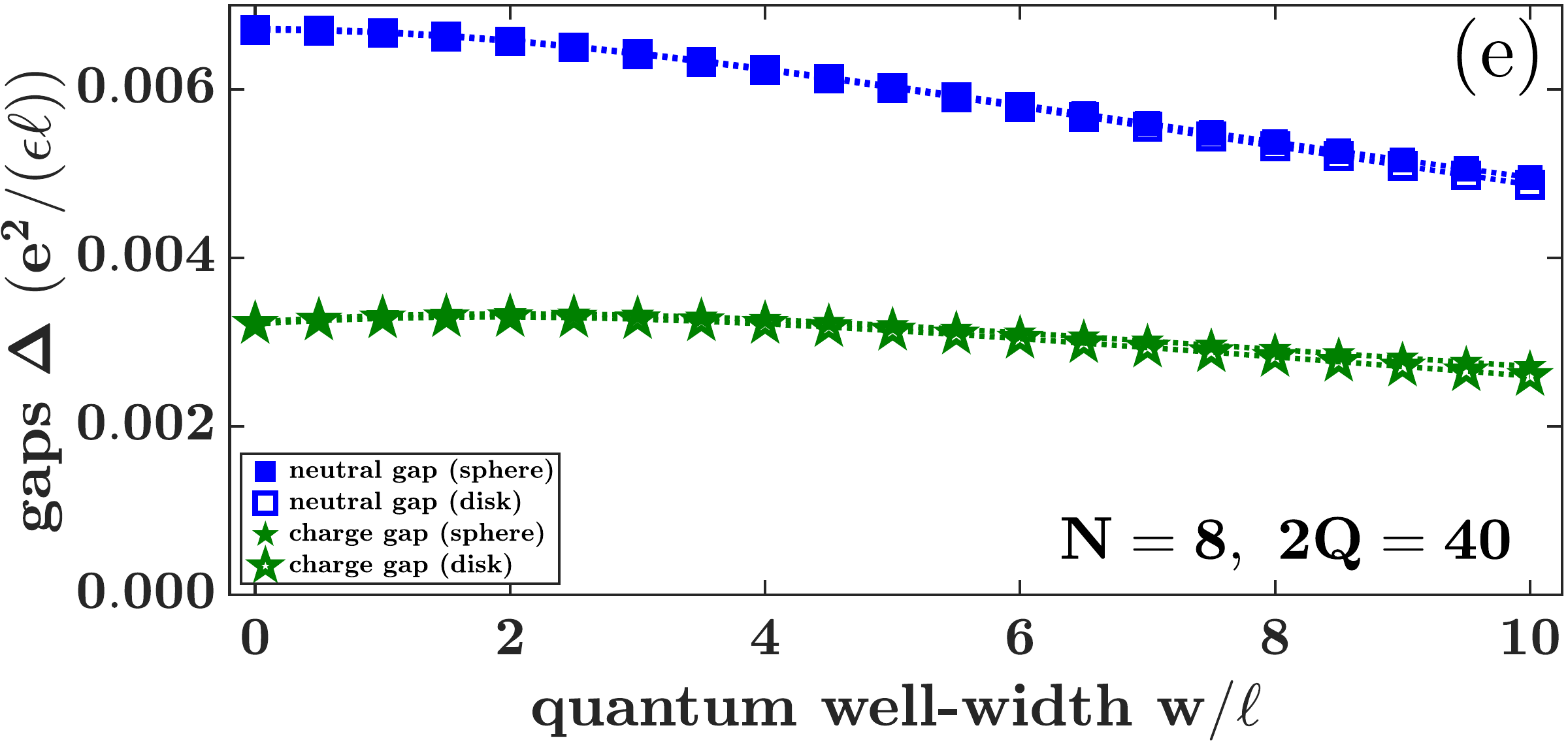} 
			\includegraphics[width=0.32\textwidth,height=0.17\textwidth]{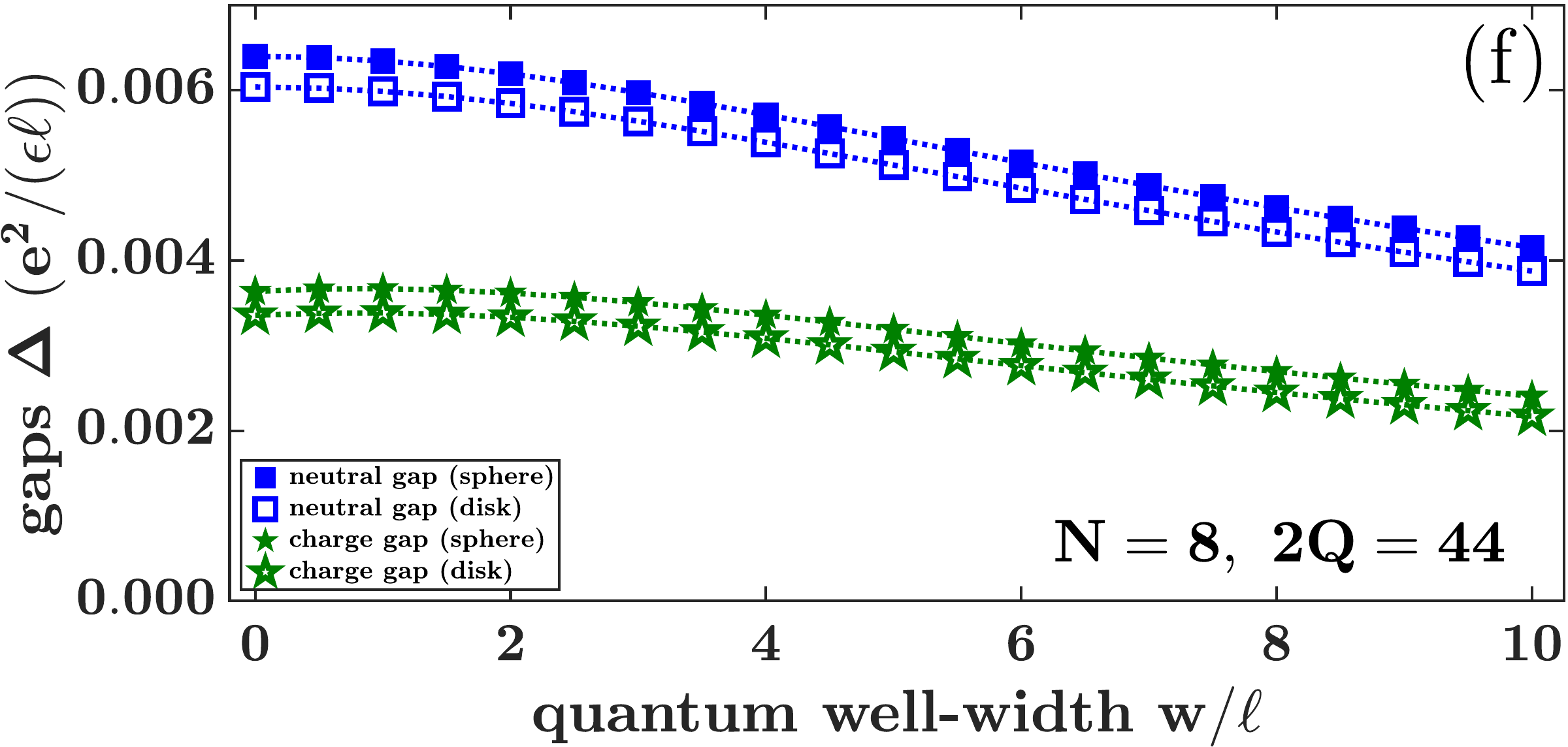} 
			\caption{(color online) Overlaps of the Laughlin-$1/7$ (top left), Jain-$2/11$ (top middle), and Jain-$2/13$ (top right) states (top panels) with the corresponding exact lowest Landau level Coulomb ground states as a function of the well-width $w/\ell$, where the width of the quantum well is modeled by a sine wave function. The lower panels show the charge and neutral gaps as a function of the well-width $w/\ell$. Results are shown for the spherical (filled symbols) and disk (open symbols) pseudopotentials. The system contains $N{=}8$ electrons.}
			\label{fig: gaps_overlaps_Laughlin_Jain_states}
		\end{center}
	\end{figure*}

\begin{figure*}[htpb]
		\begin{center}
			\includegraphics[width=0.32\textwidth,height=0.17\textwidth]{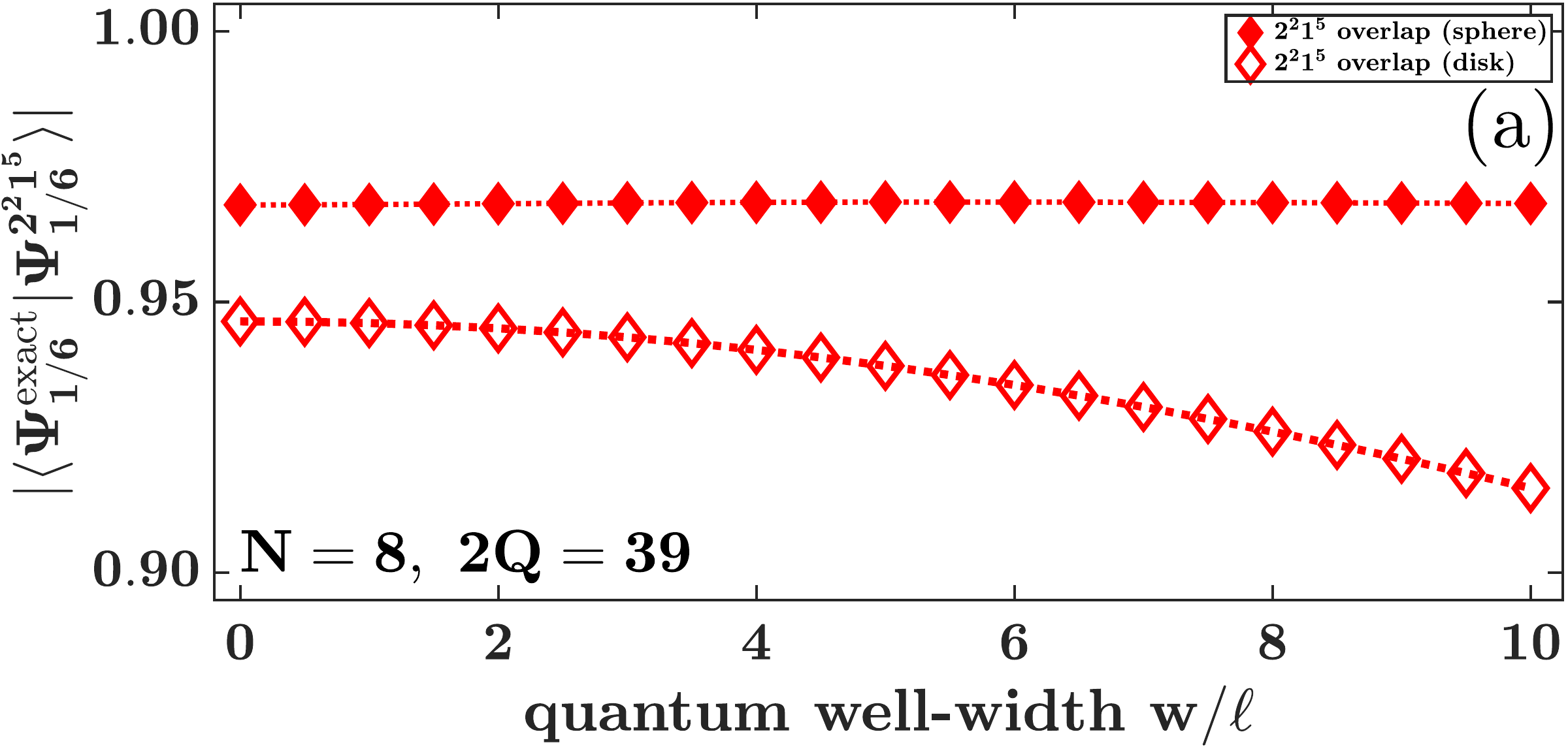}		
			\includegraphics[width=0.32\textwidth,height=0.17\textwidth]{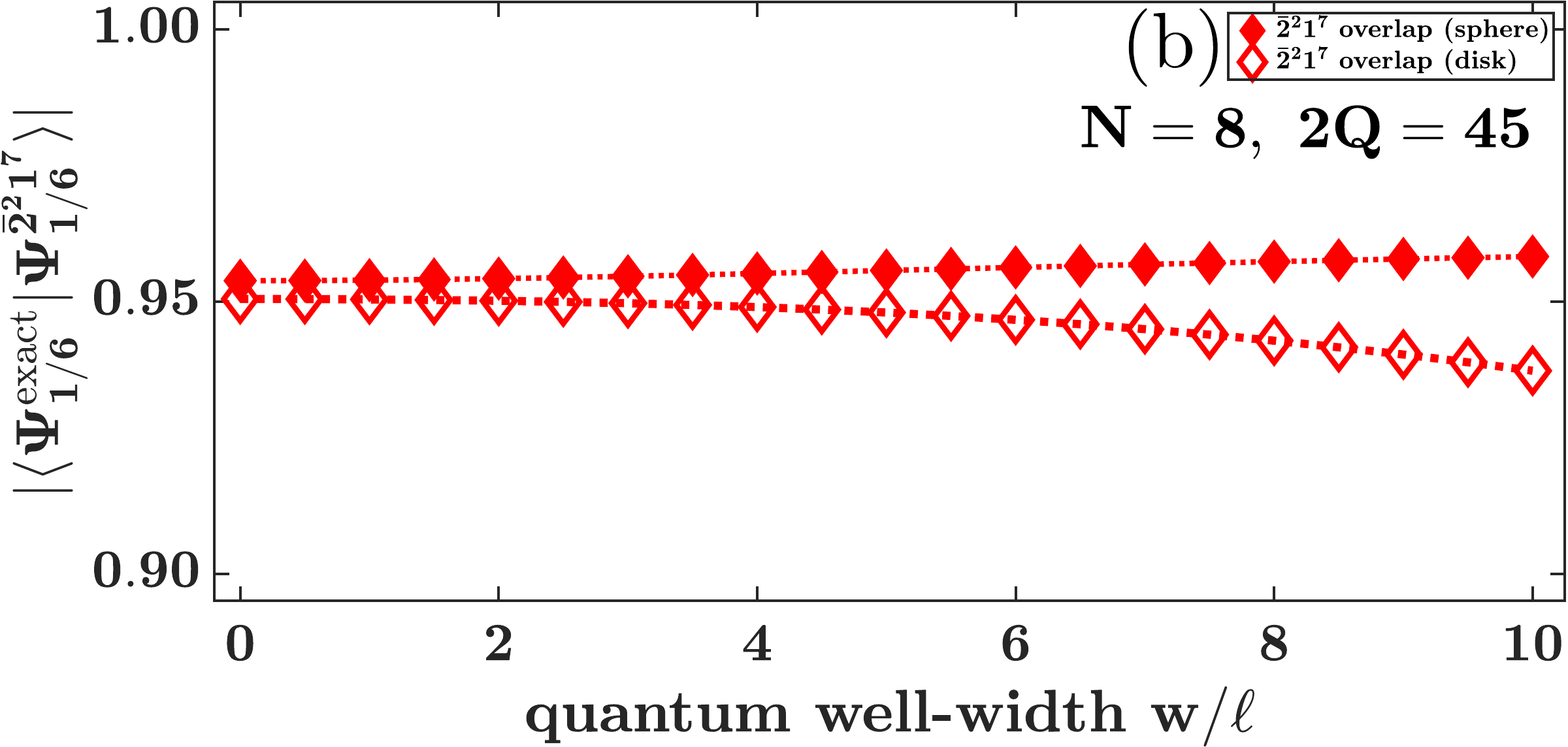} 
			\includegraphics[width=0.32\textwidth,height=0.17\textwidth]{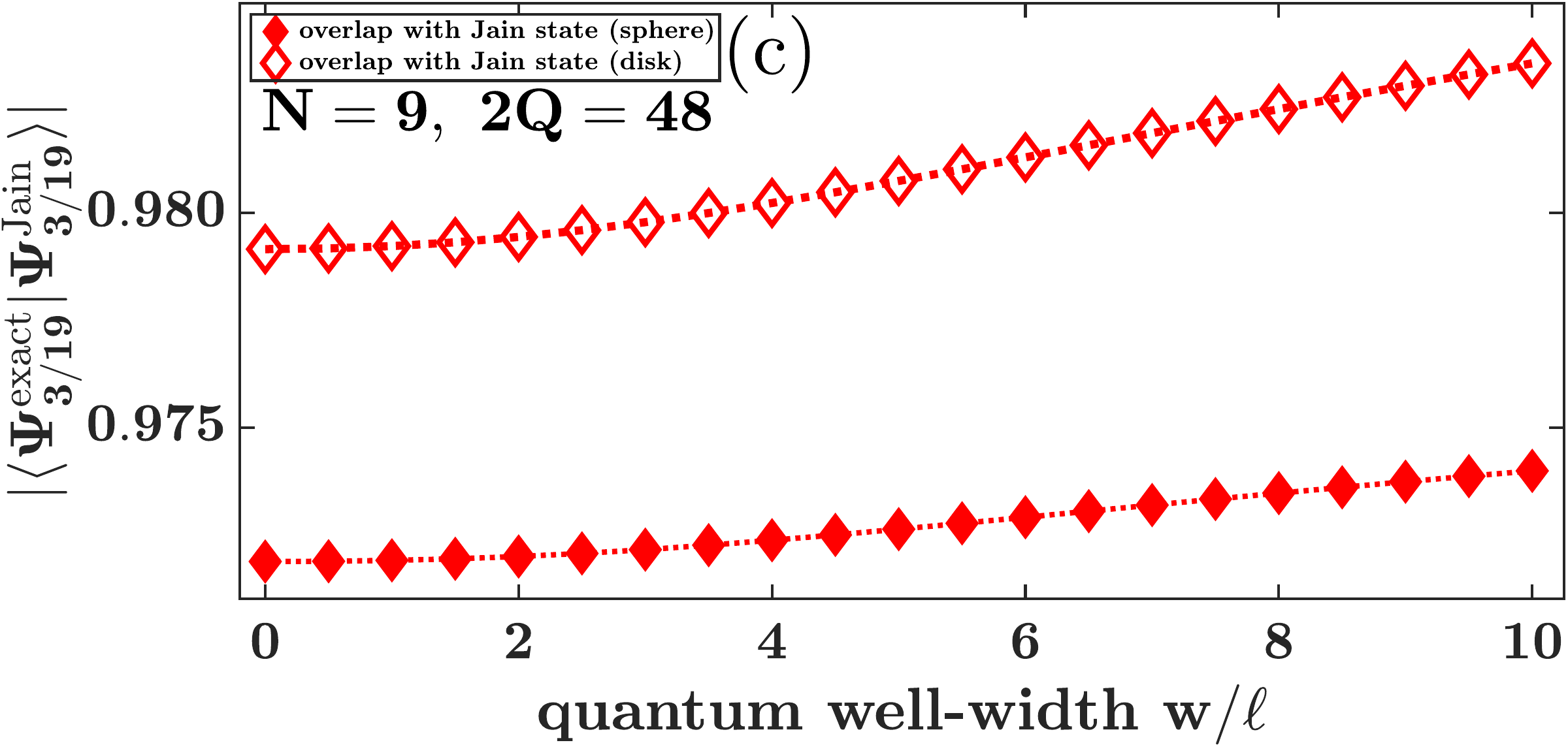} 
			\\
			\includegraphics[width=0.32\textwidth,height=0.17\textwidth]{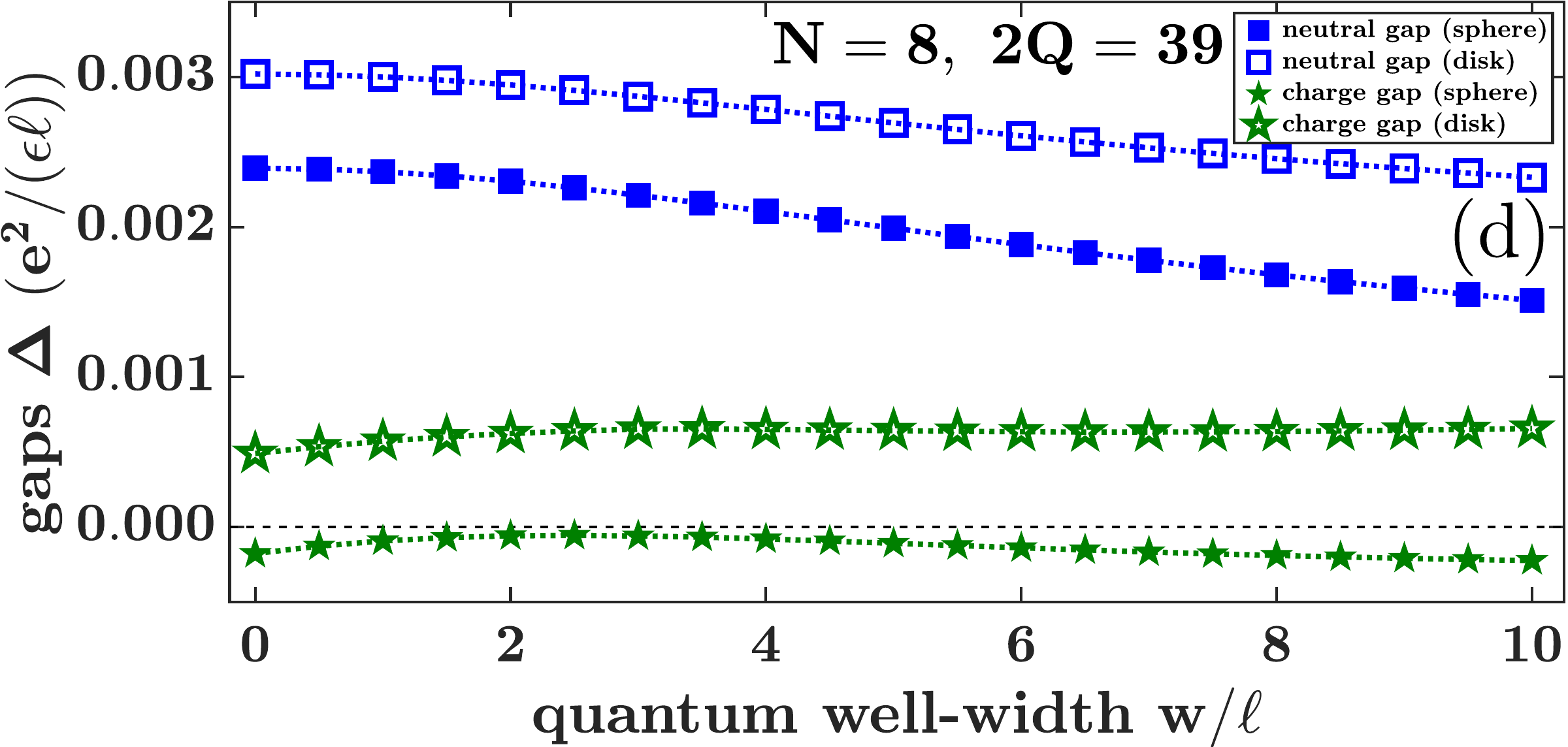} 	
			\includegraphics[width=0.32\textwidth,height=0.17\textwidth]{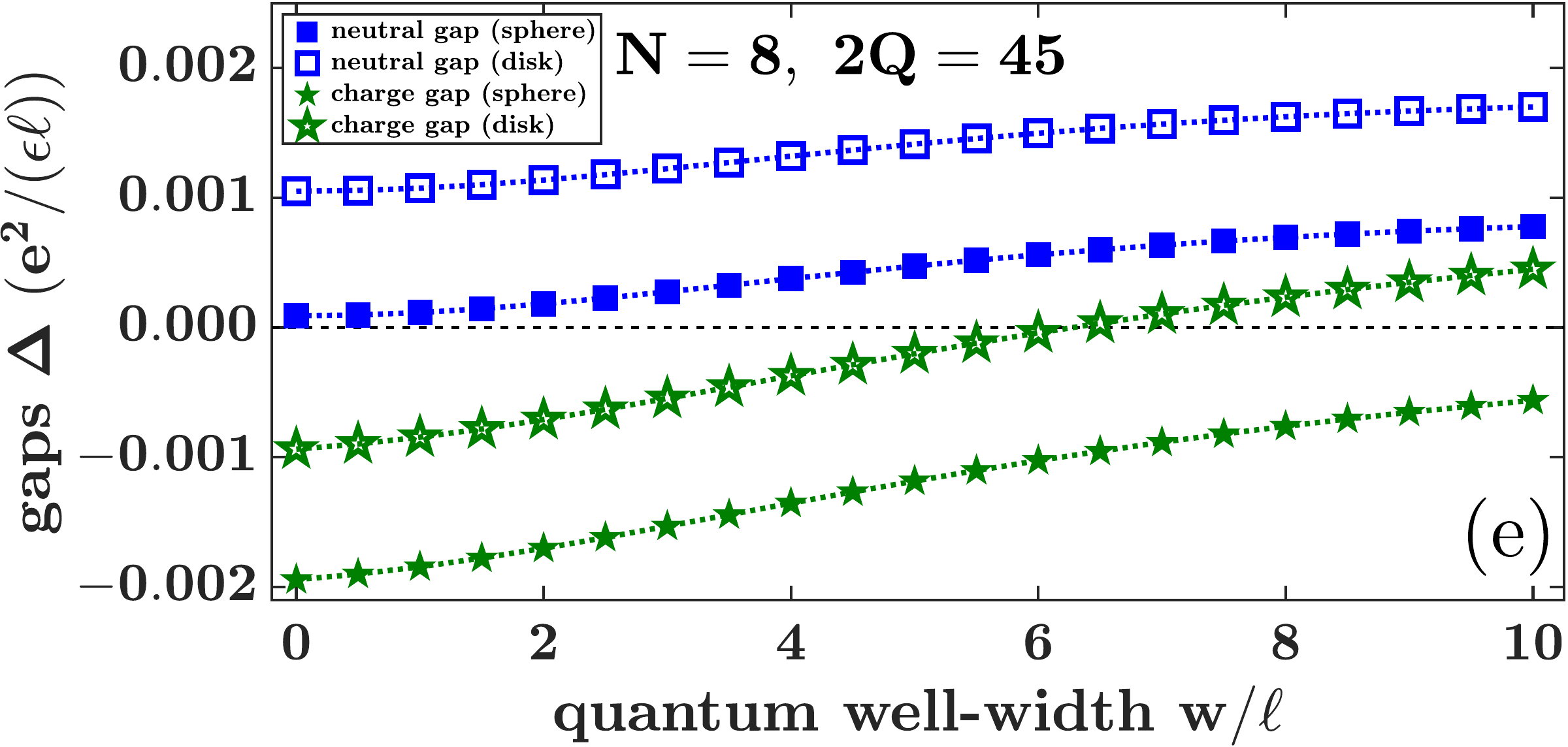} 
			\includegraphics[width=0.32\textwidth,height=0.17\textwidth]{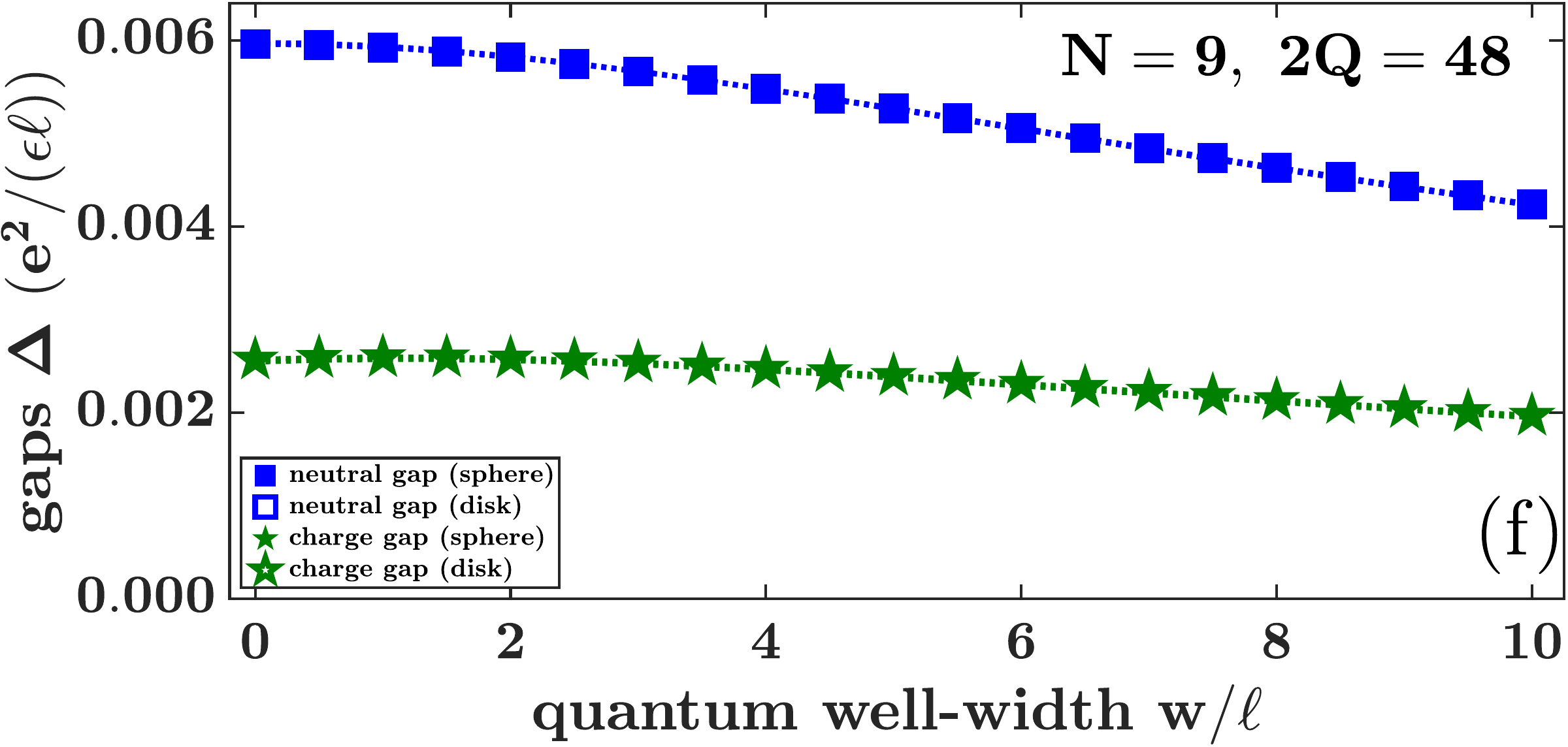} 
			\caption{(color online) Overlaps (top panels) of the $2^{2}1^{5}$-parton (left panels), $\bar{2}^{2}1^{7}$-parton (middle panels), and CF Fermi sea states (right panels) with the exact lowest Landau level Coulomb ground state and charge and neutral gaps (bottom panels) as a function of the well-width $w/\ell$ at $\nu{=}1/6$. Results are shown for the spherical (filled symbols) and disk (open symbols) pseudopotentials, obtained by assuming a sine wave function in the transverse direction. The $N{=}9$ filled-shell CF Fermi sea system at $\nu{=}1/6$ aliases with the Jain states at $\nu{=}3/19$ and $\nu{=}3/17$, and the charge gaps are calculated assuming the system forms an incompressible state at these Jain fractions.}
			\label{fig: gaps_overlaps_candidate_states_1_6}
		\end{center}
	\end{figure*}

    In Figs.~\ref{fig: gaps_overlaps_Laughlin_Jain_states} and \ref{fig: gaps_overlaps_candidate_states_1_6} we show the overlaps and gaps of the Laughlin, Jain, and candidate states at $\nu{=}1/6$ respectively, for a finite-width system where the width of the quantum well is modeled by a sine wave function. We compute the ED ground states using spherical and planar disk pseudopotentials. The overlaps of the $1/7$ Laughlin, $2/11$, $2/13$, and $3/19$ Jain, $2^{2}1^{5}$ and $\bar{2}^{2}1^{7}$ parton states with the ED ground states are fairly high up to widths of $w{=}10\ell$. Furthermore, the gaps for the Laughlin and Jain states are quite robust, but that is not the case for the parton states. In certain cases, the neutral gap is lower than the charge gap, which indicates strong finite-size effects since in the thermodynamic limit the charge gap is at least as large as the neutral gap~\cite{Lemm24}. Moreover, on the sphere, there could be possible aliasing issues~\cite{Ambrumenil88} since the same system could be a finite-size representation of thermodynamic states at two different fillings. For example, the charge gap of the $2^{2}1^{5}$ parton state requires considering the state obtained by adding one flux to it. For $N{=}8$ particles, this flux-added state occurs at $N{=}8,~2Q{=}40$. However, the $N{=}8,~2Q{=}40$ is also a finite size representation of the $2/11$ Jain state. This could be why the charge gap for the parton states for some parameters is negative. Generically, we find that the $2^{2}1^{5}$ state has higher overlaps and gaps compared to the $\bar{2}^{2}1^{7}$ state. The ground state at the Moore-Read Pfaffian flux does not have $L{=}0$ for $N{=}8$ for $0{\leq}w{\leq}10\ell$.   

    %%%%%%%%%%%%%%%%%%%%%%%%%%%%%%%%%%%%%%%%%%%%%%%%%%%%%%%%%%%%%%%%%%%%%%%%%%%%%
	%%%%%%%%%%%%%%%%%%%%%%%%%%%%%%%%%%%%%%%%%%%%%%%%%%%%%%%%%%%%%%%%%%%%%%%%%%%%%

	%%%%%%%%%%%%%%%%%%%%%%%%%%%%%%%%%%%%%%%%%%%%%%%%%%%%%%%%%%%%%%%%%%%%%%%%%%%%%
	%%%%%%%%%%%%%%%%%%%%%%%%%%%%%%%%%%%%%%%%%%%%%%%%%%%%%%%%%%%%%%%%%%%%%%%%%%%%%
    
\begin{figure*}[htpb]
		\begin{center}			
			\includegraphics[width=0.32\textwidth]{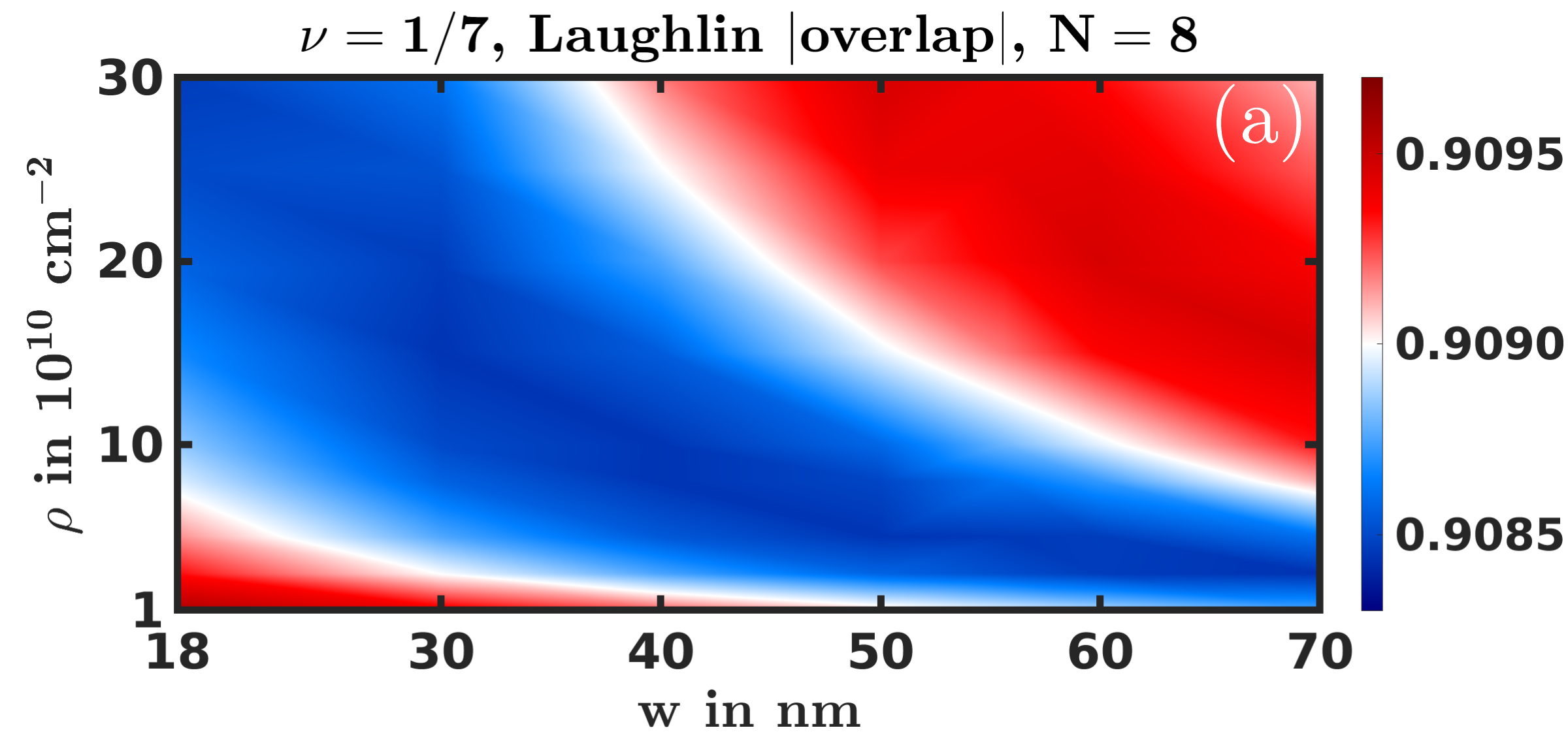}
			\includegraphics[width=0.32\textwidth]{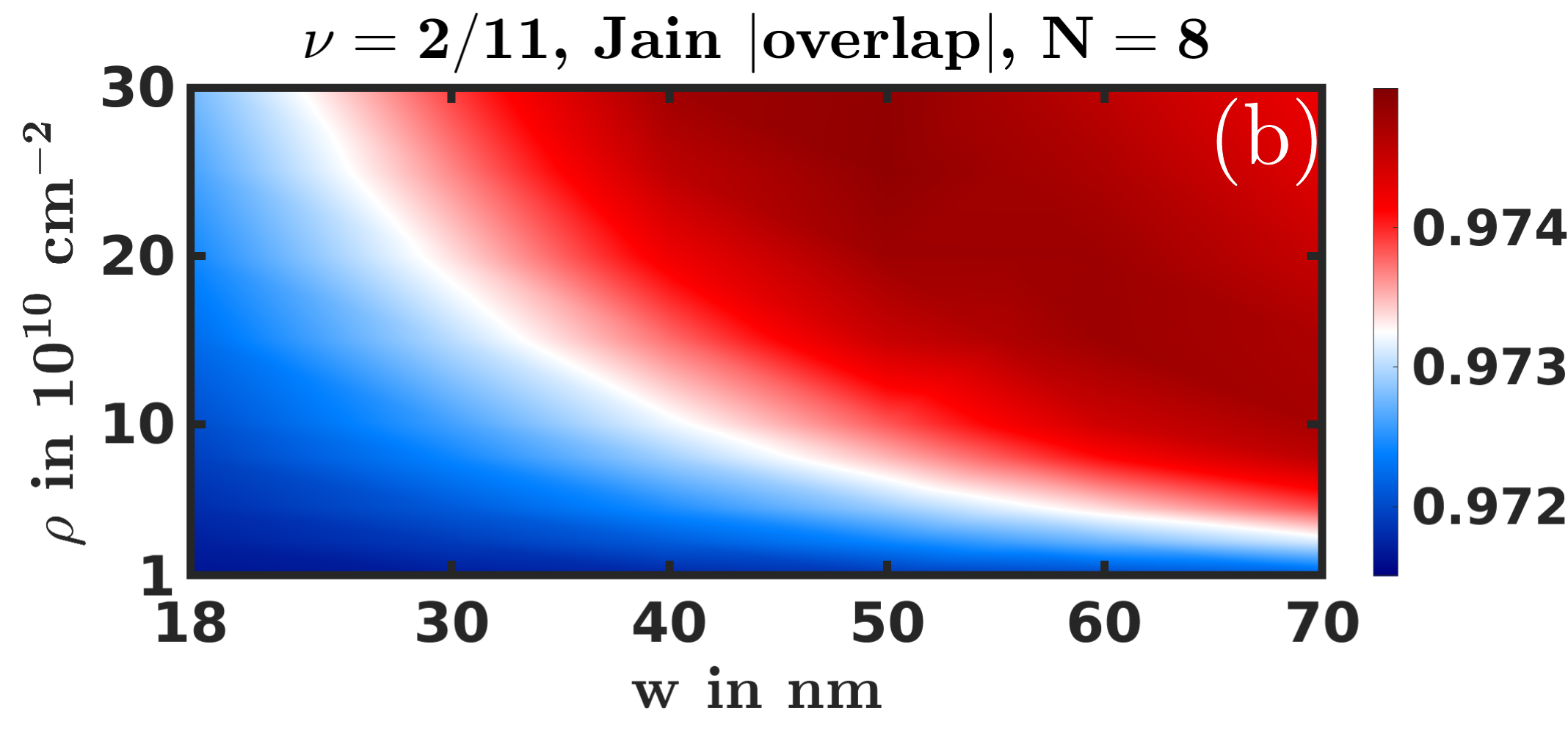}
			\includegraphics[width=0.32\textwidth]{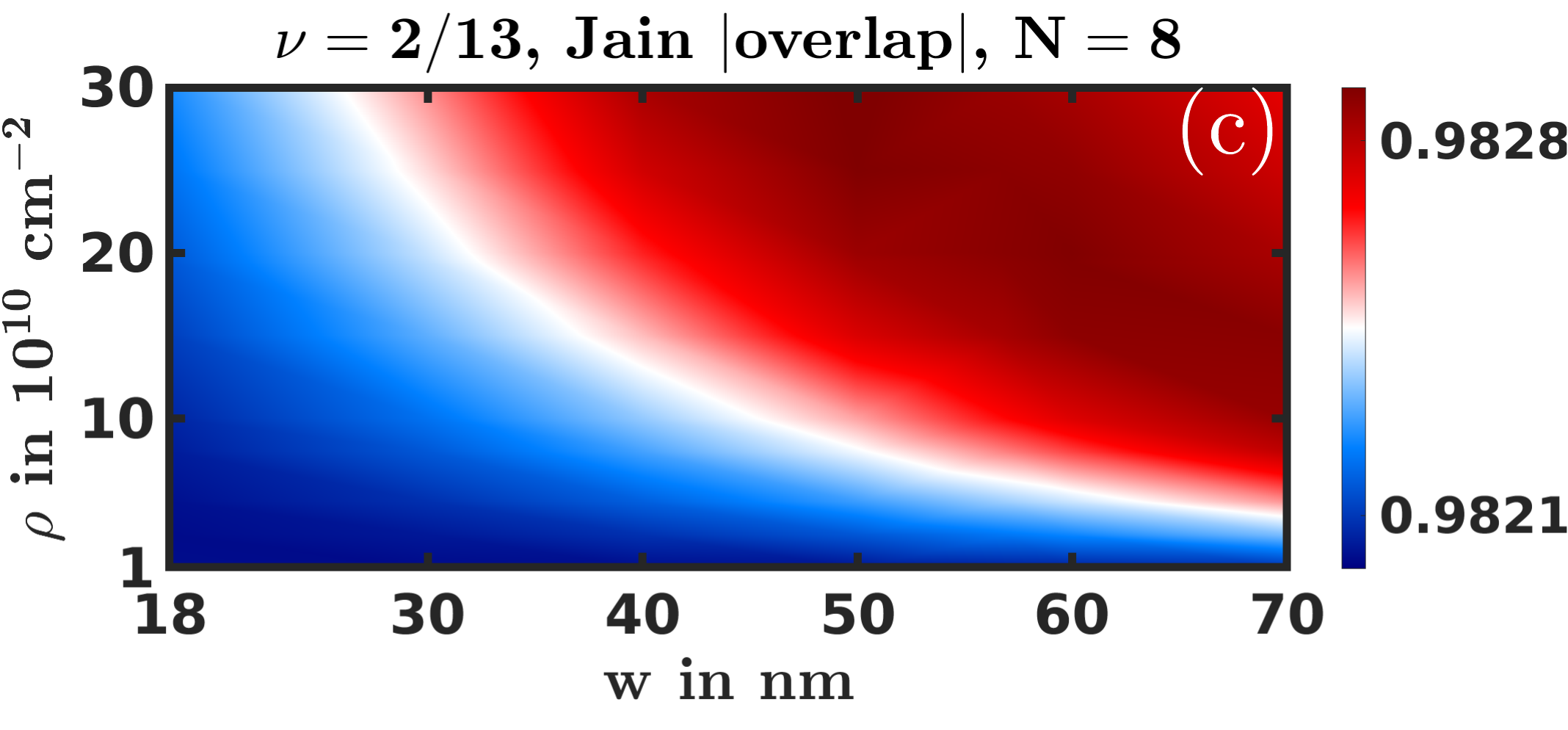}\\
			\vspace{0.3cm}
			\includegraphics[width=0.32\textwidth]{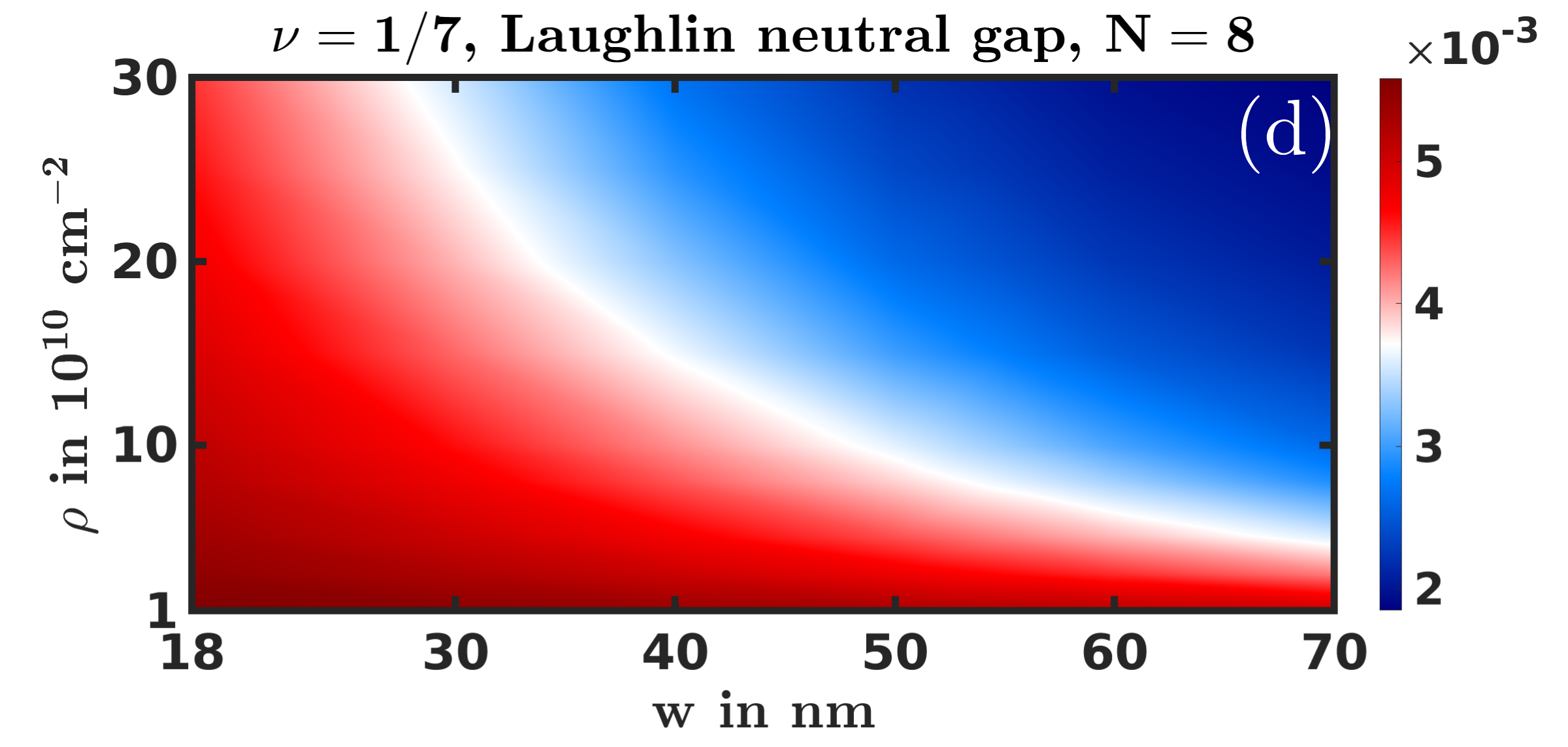}
			\includegraphics[width=0.32\textwidth]{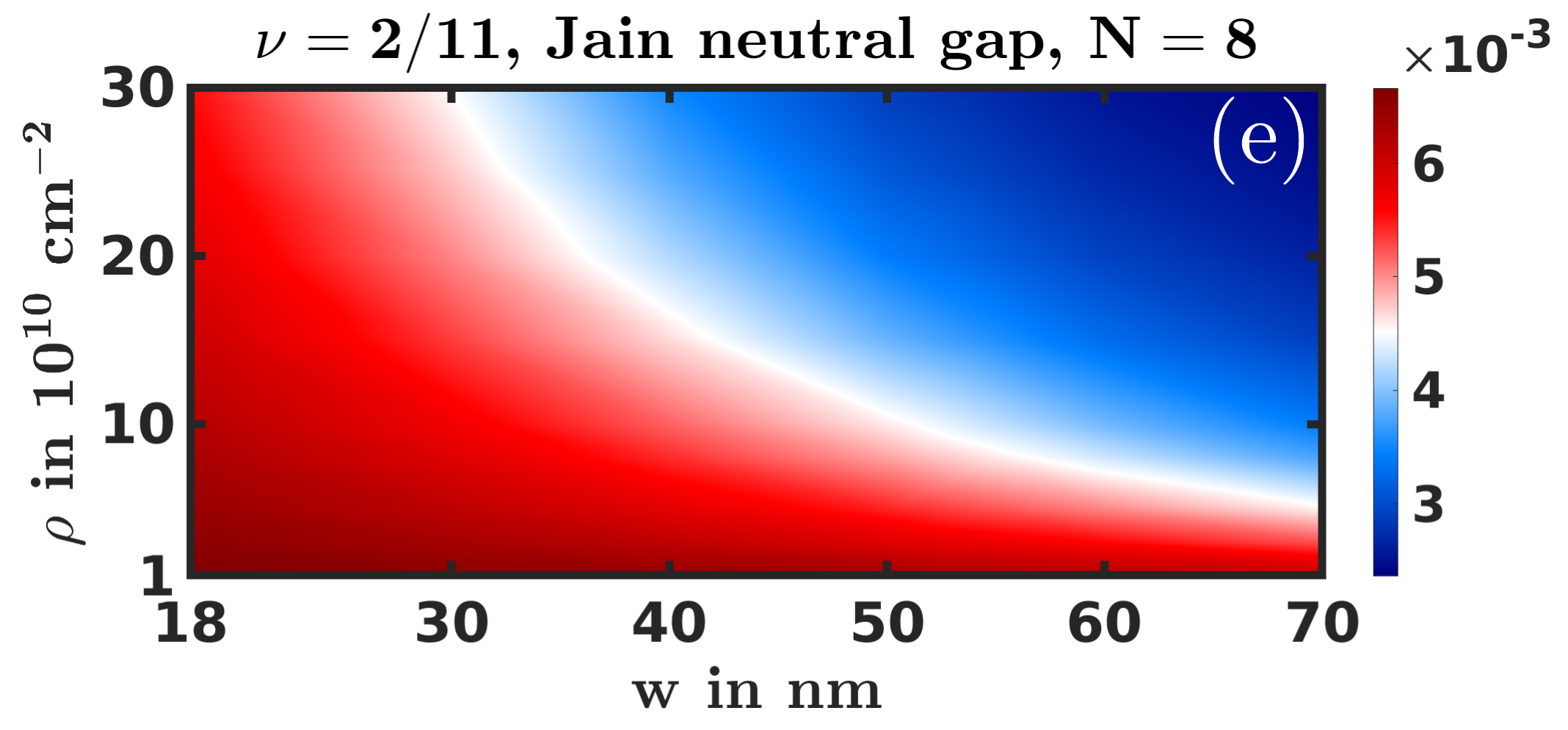}
			\includegraphics[width=0.32\textwidth]{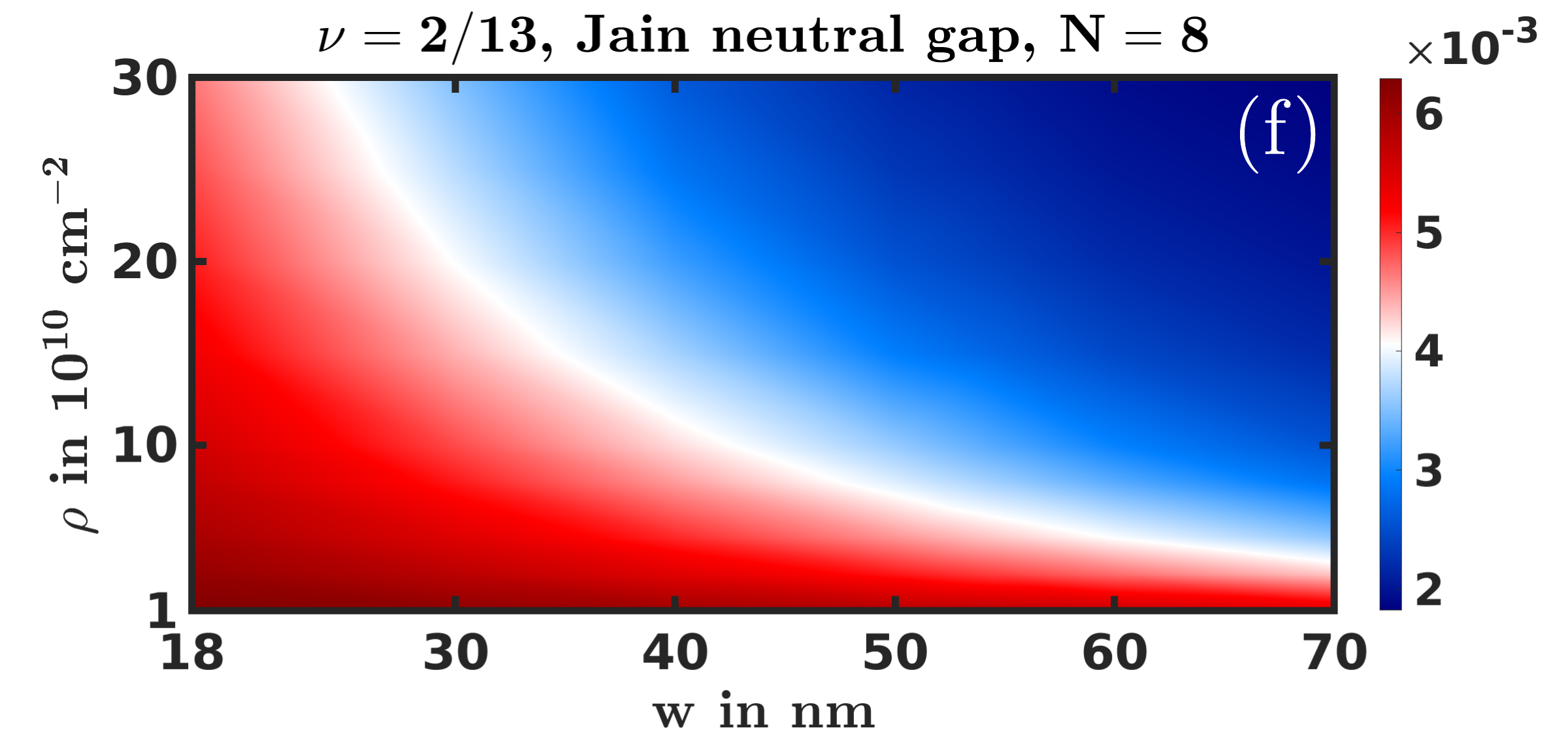} \\
			\vspace{0.3cm}
			\includegraphics[width=0.32\textwidth]{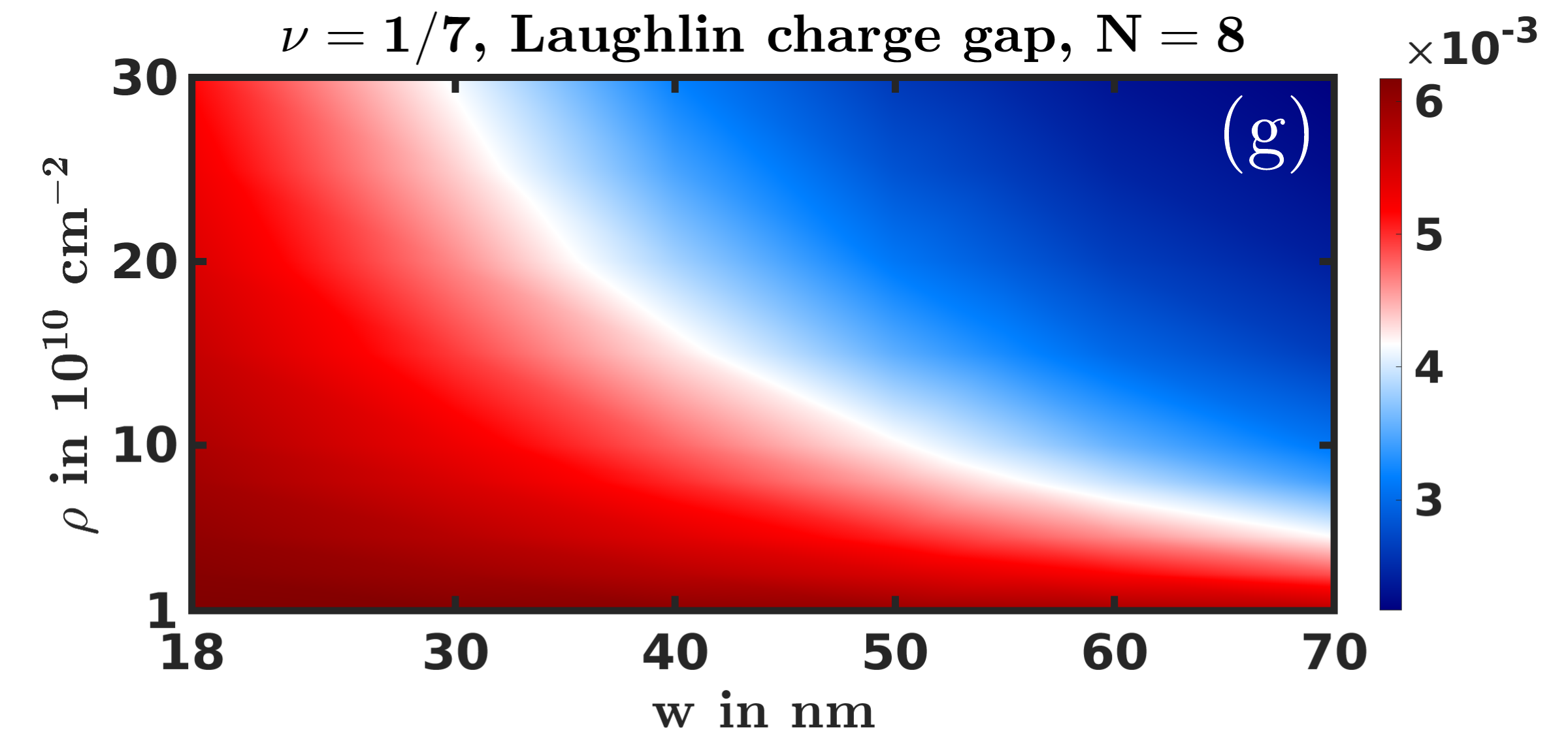}
			\includegraphics[width=0.32\textwidth]{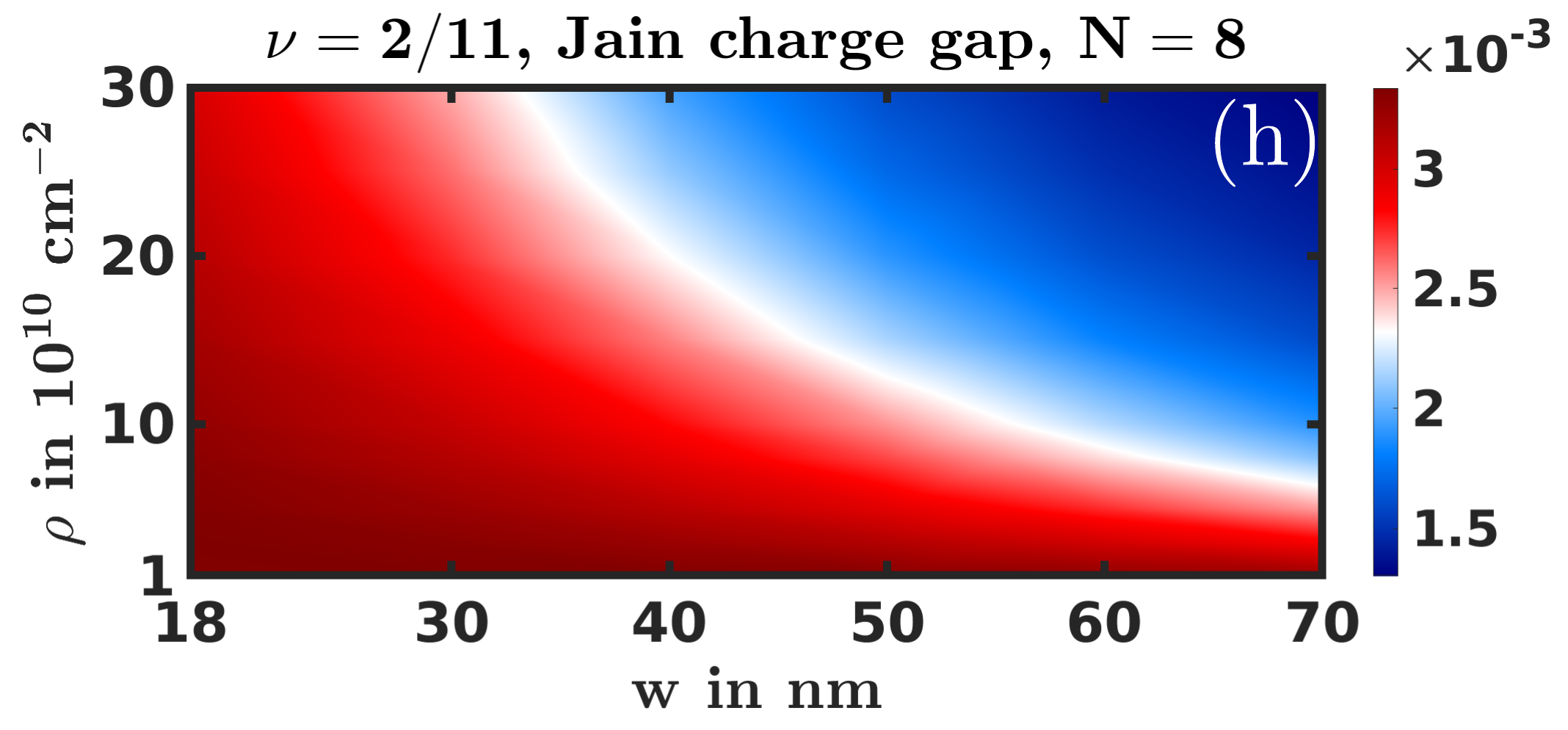}
			\includegraphics[width=0.32\textwidth]{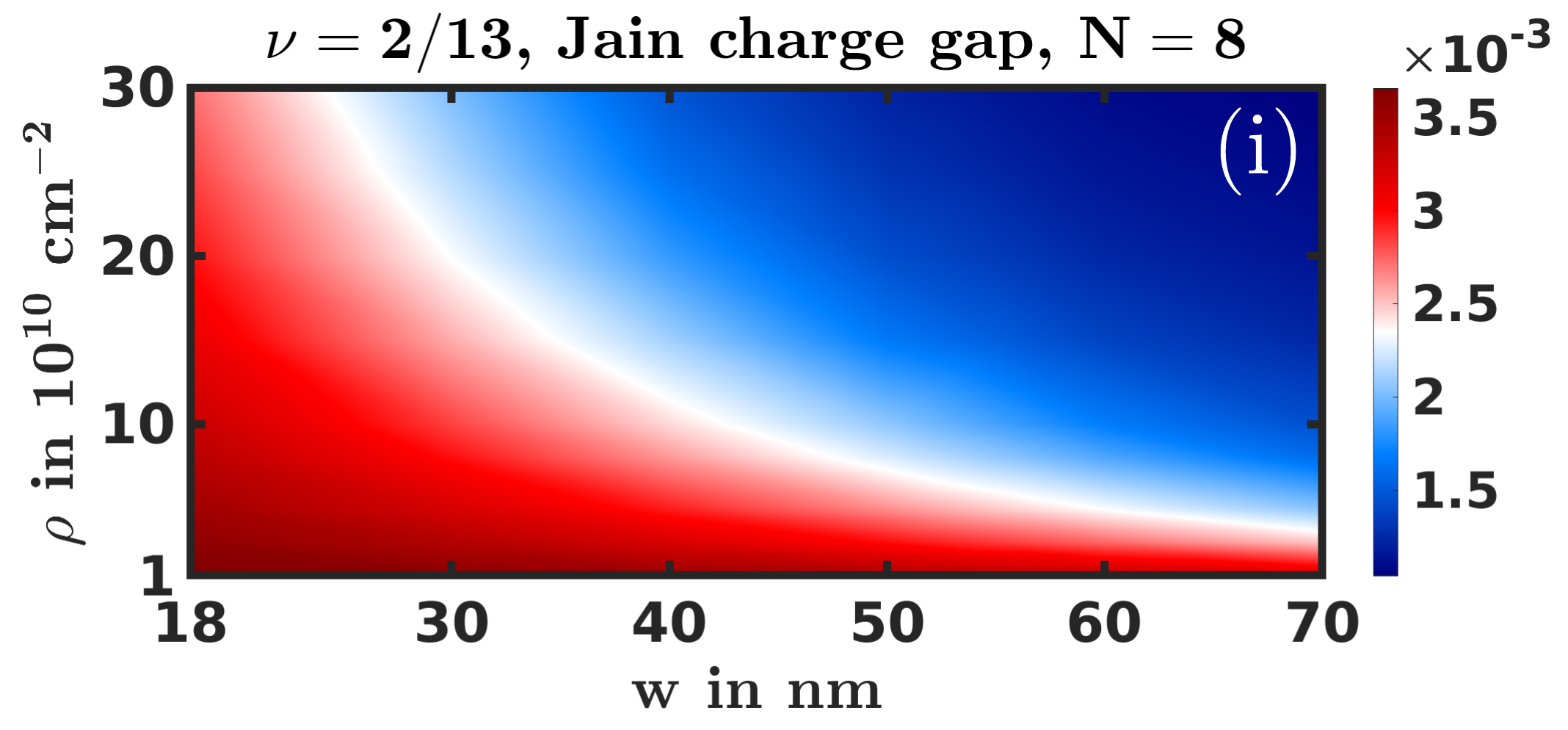}
		\end{center}
		\caption{
			Panels (a), (b), and (c) show the overlaps of the Laughlin-$1/7$, Jain-$2/11$, and Jain-$2/13$  states with the corresponding exact lowest Landau level ground states as a function of the electron density and the quantum-well width, using the pseudopotentials of the finite-width interaction obtained using a local density approximation (LDA). The panels (d), (e), and (f) show the corresponding neutral gaps, and (g), (h), and (i) the corresponding charge gaps. All systems have $N{=}8$ electrons in the spherical geometry.}
		\label{fig: Laughlin_Jain_overlaps_gaps_finite_width_LDA}
	\end{figure*}

\begin{figure*}[htpb]
		\begin{center}			
			\includegraphics[width=0.32\textwidth]{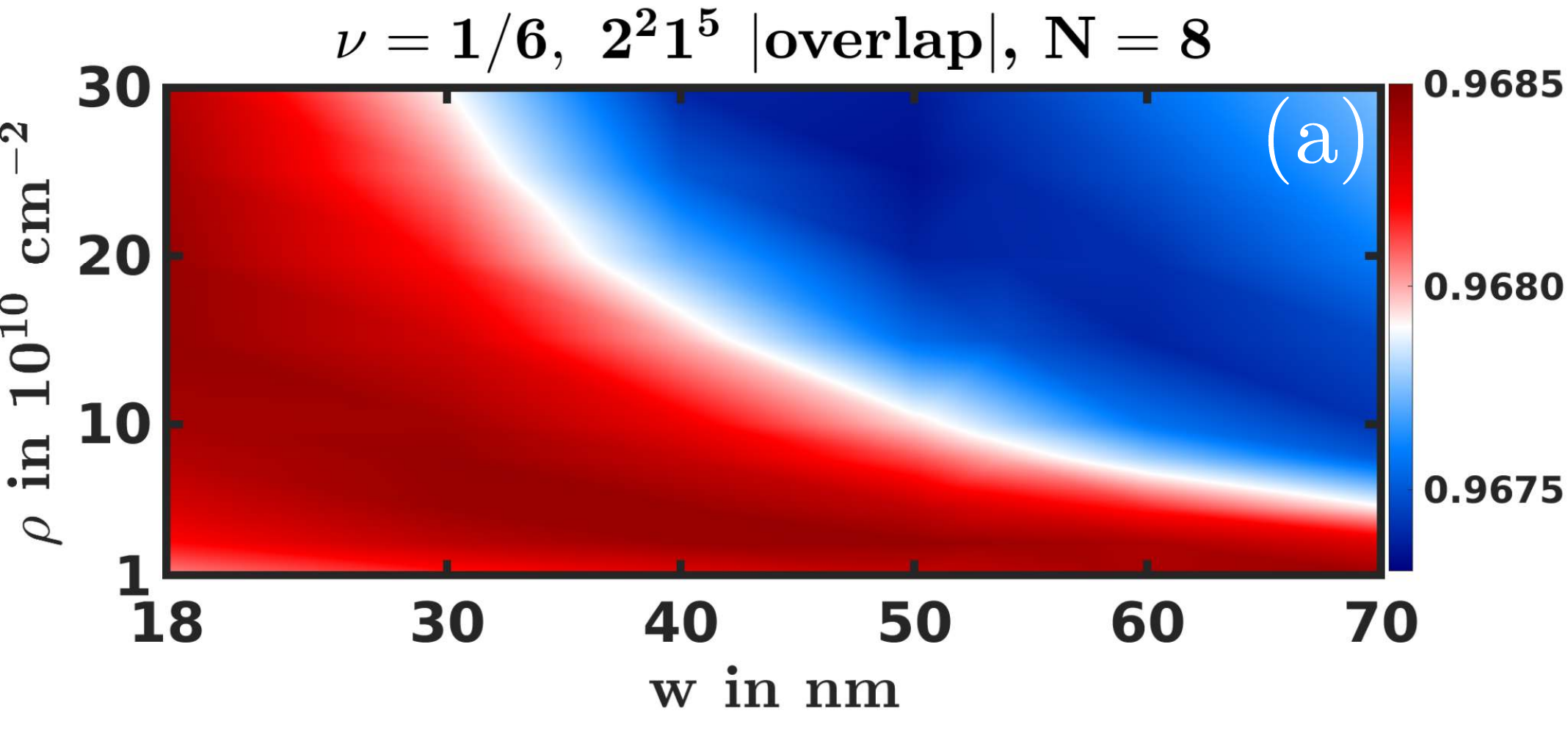}
			\includegraphics[width=0.32\textwidth]{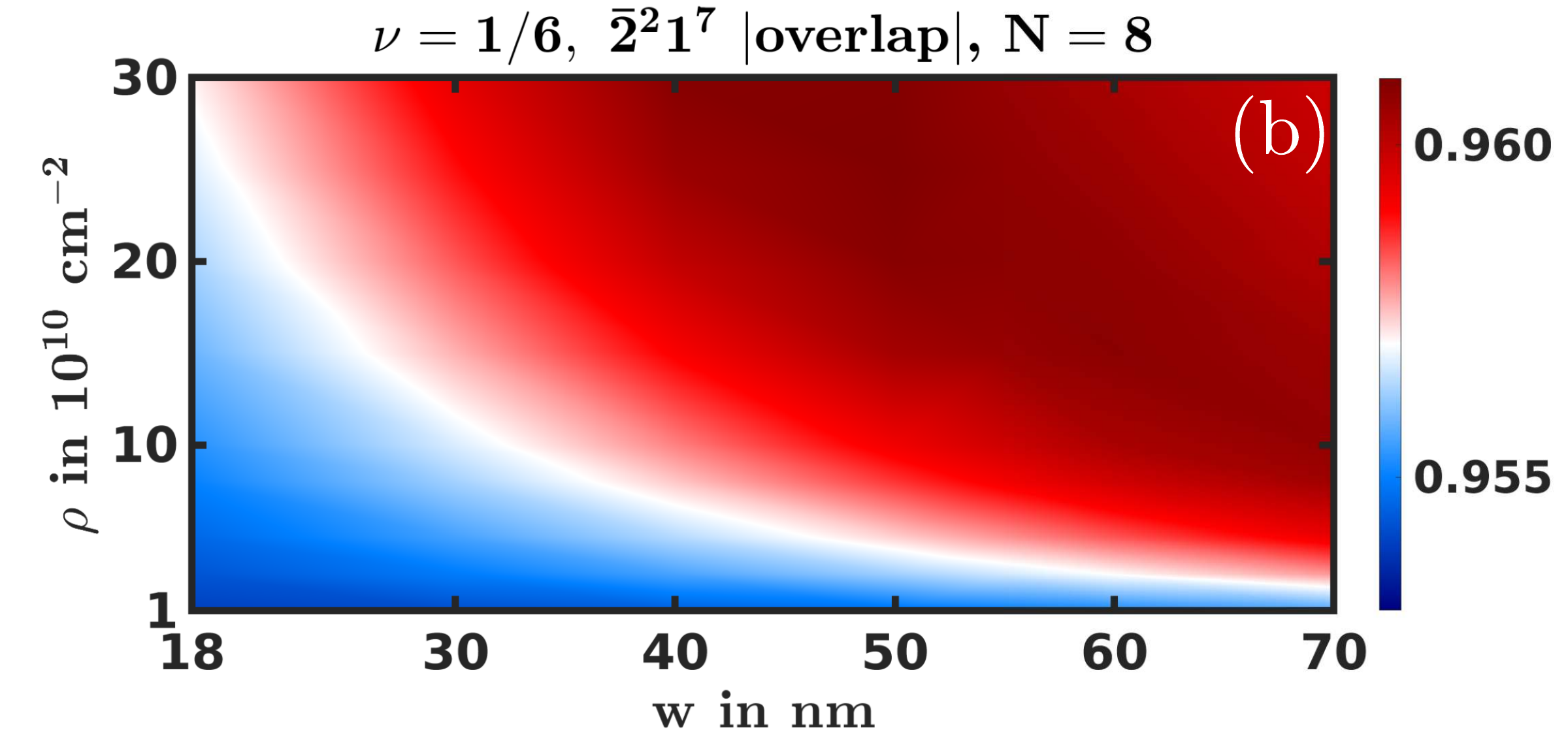}
			\includegraphics[width=0.32\textwidth]{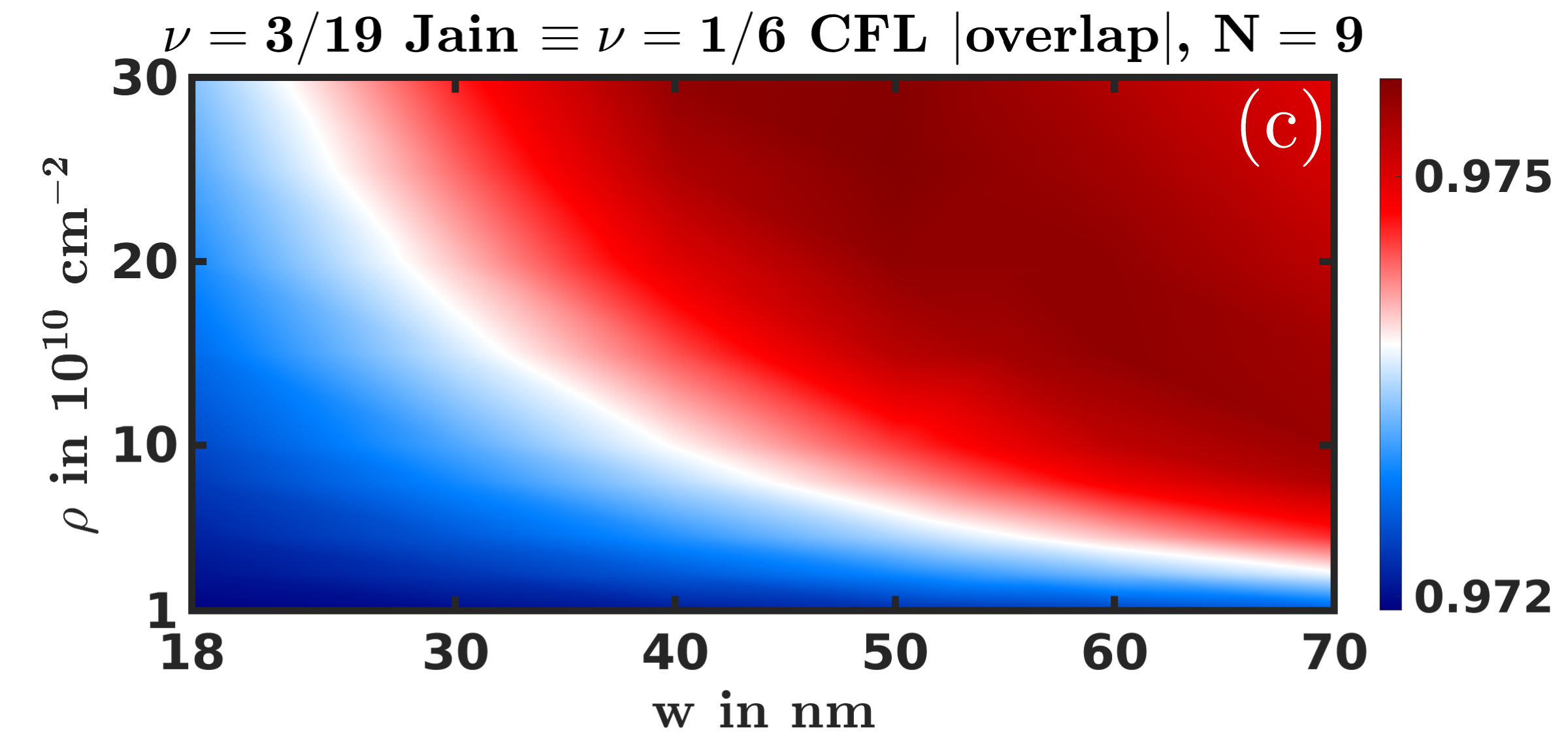}\\
			\vspace{0.3cm}
			\includegraphics[width=0.32\textwidth]{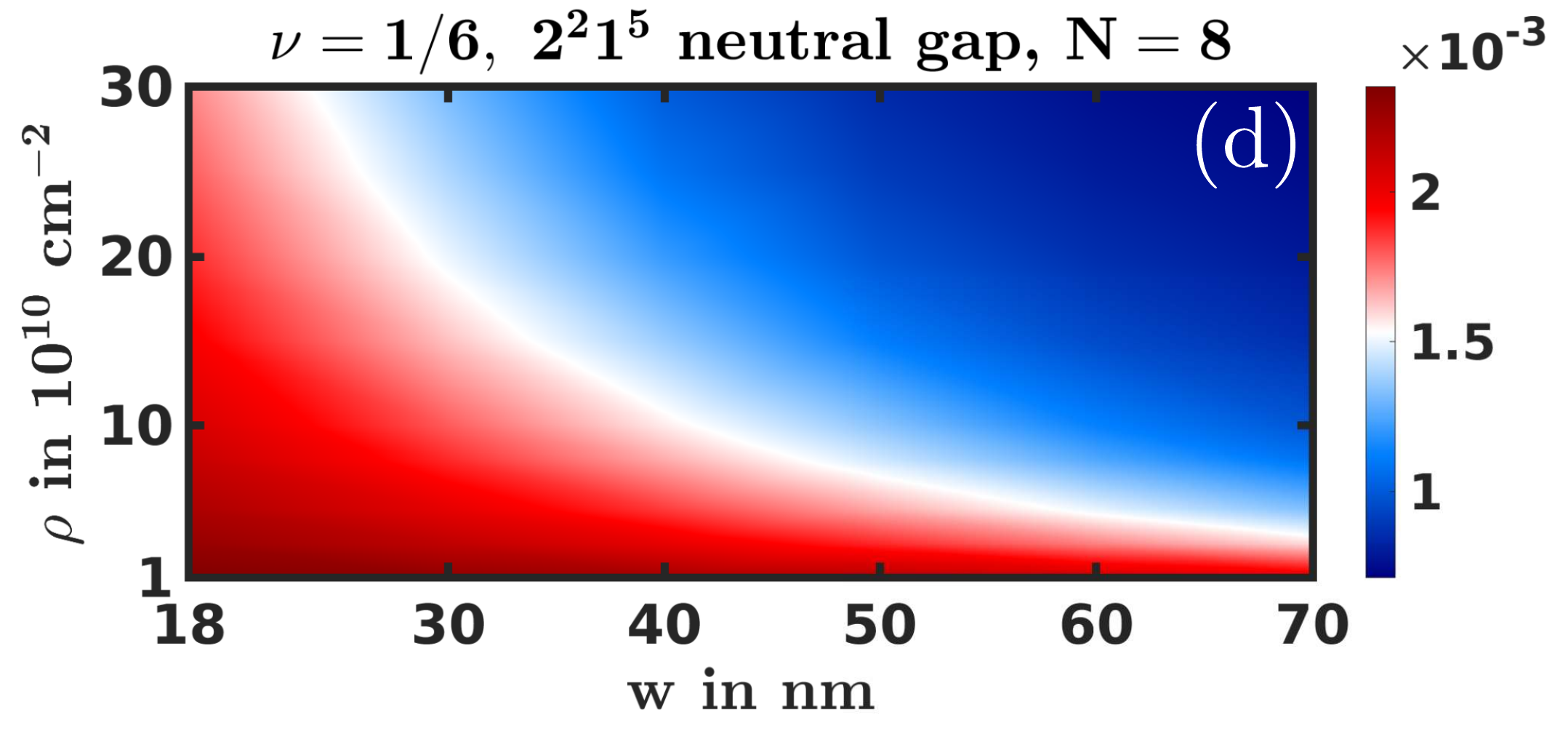}
			\includegraphics[width=0.32\textwidth]{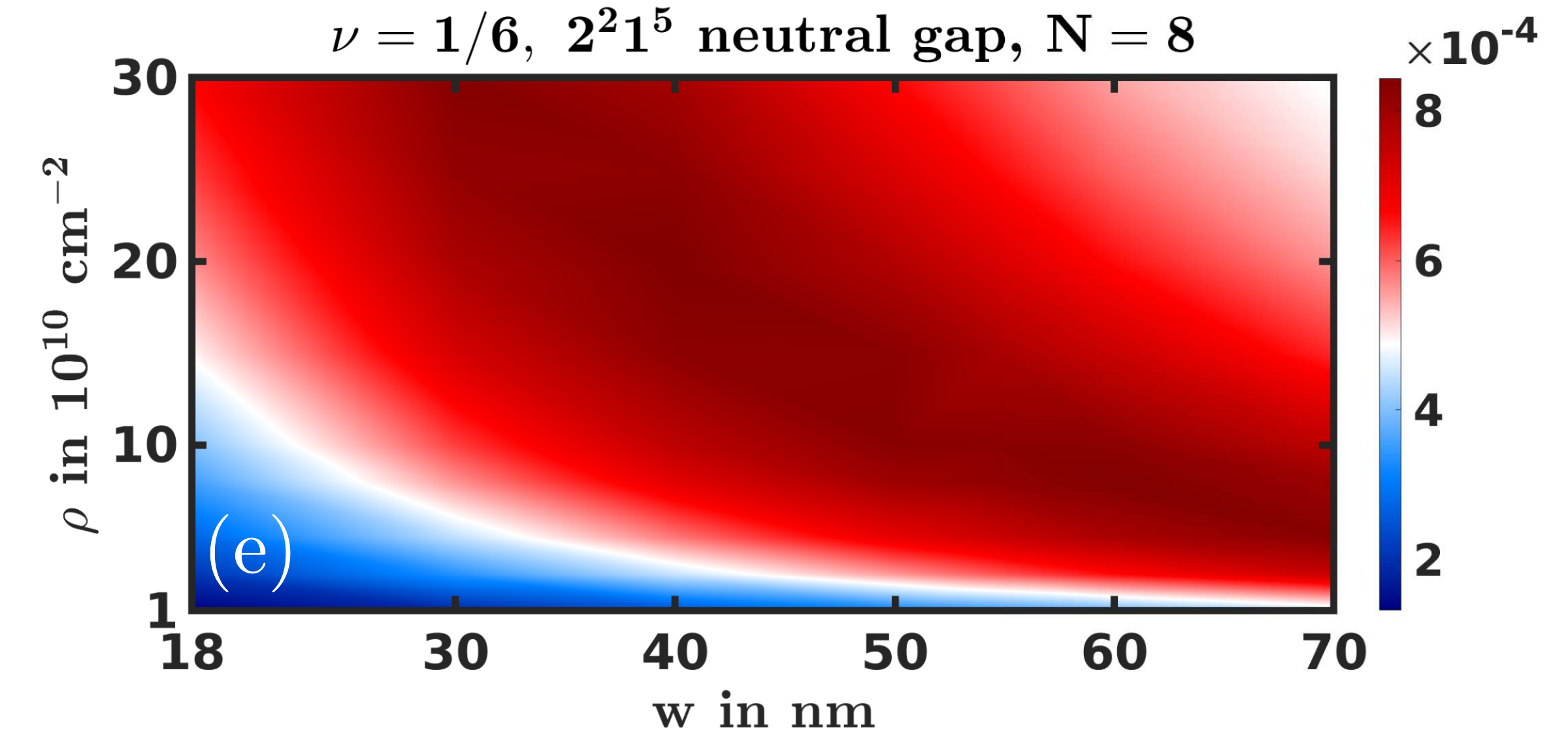}
			\includegraphics[width=0.32\textwidth]{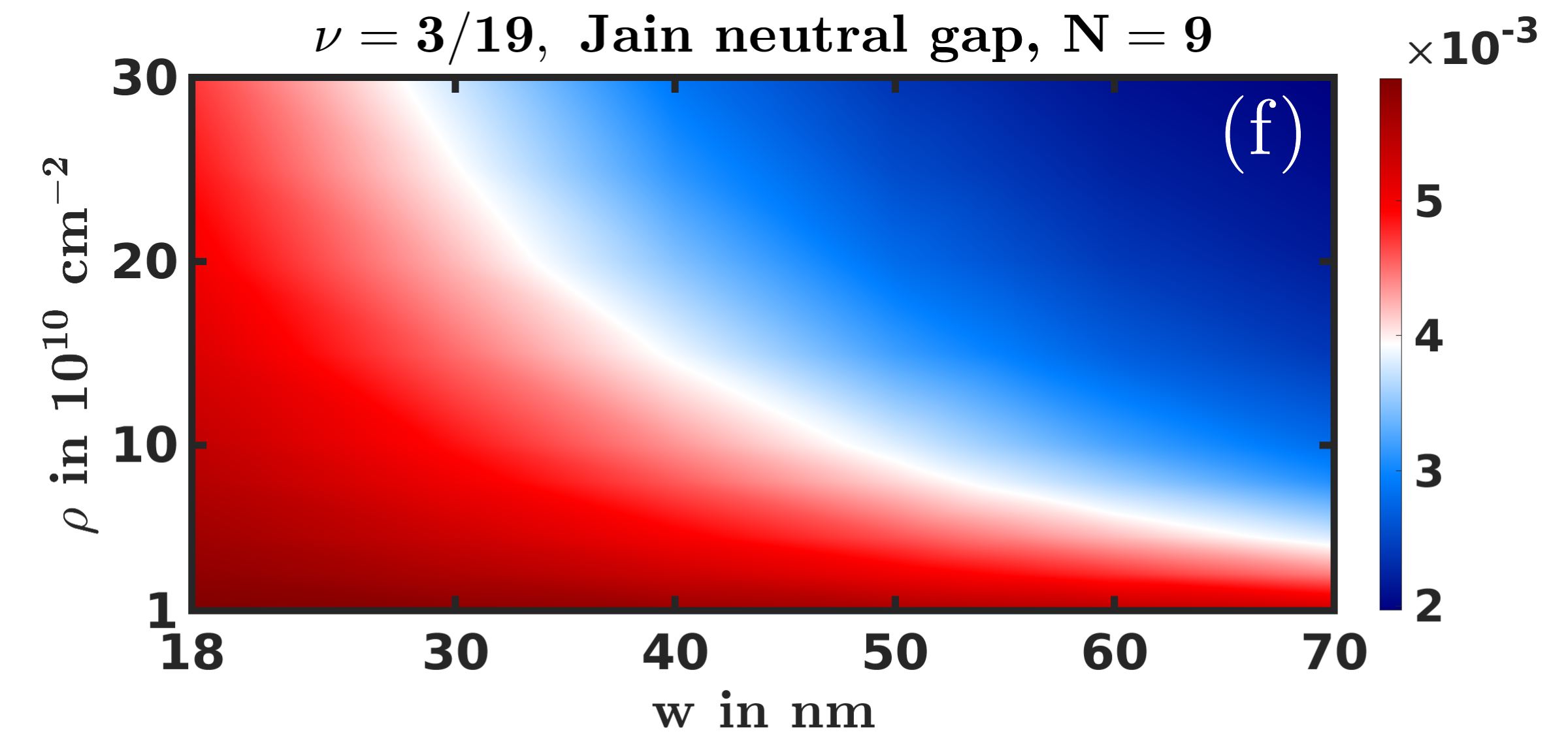} \\
			\vspace{0.3cm}
			\includegraphics[width=0.32\textwidth]{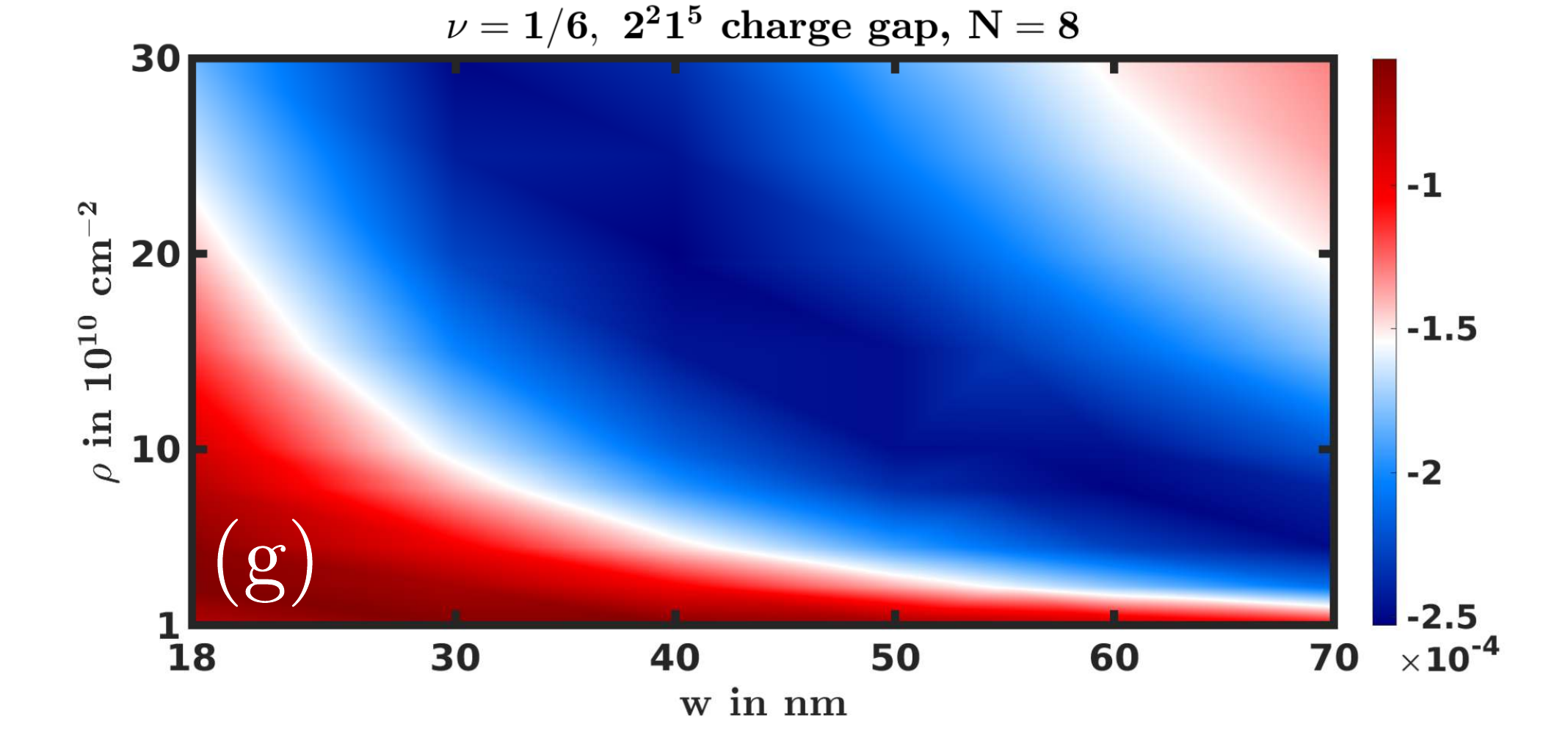}
			\includegraphics[width=0.32\textwidth]{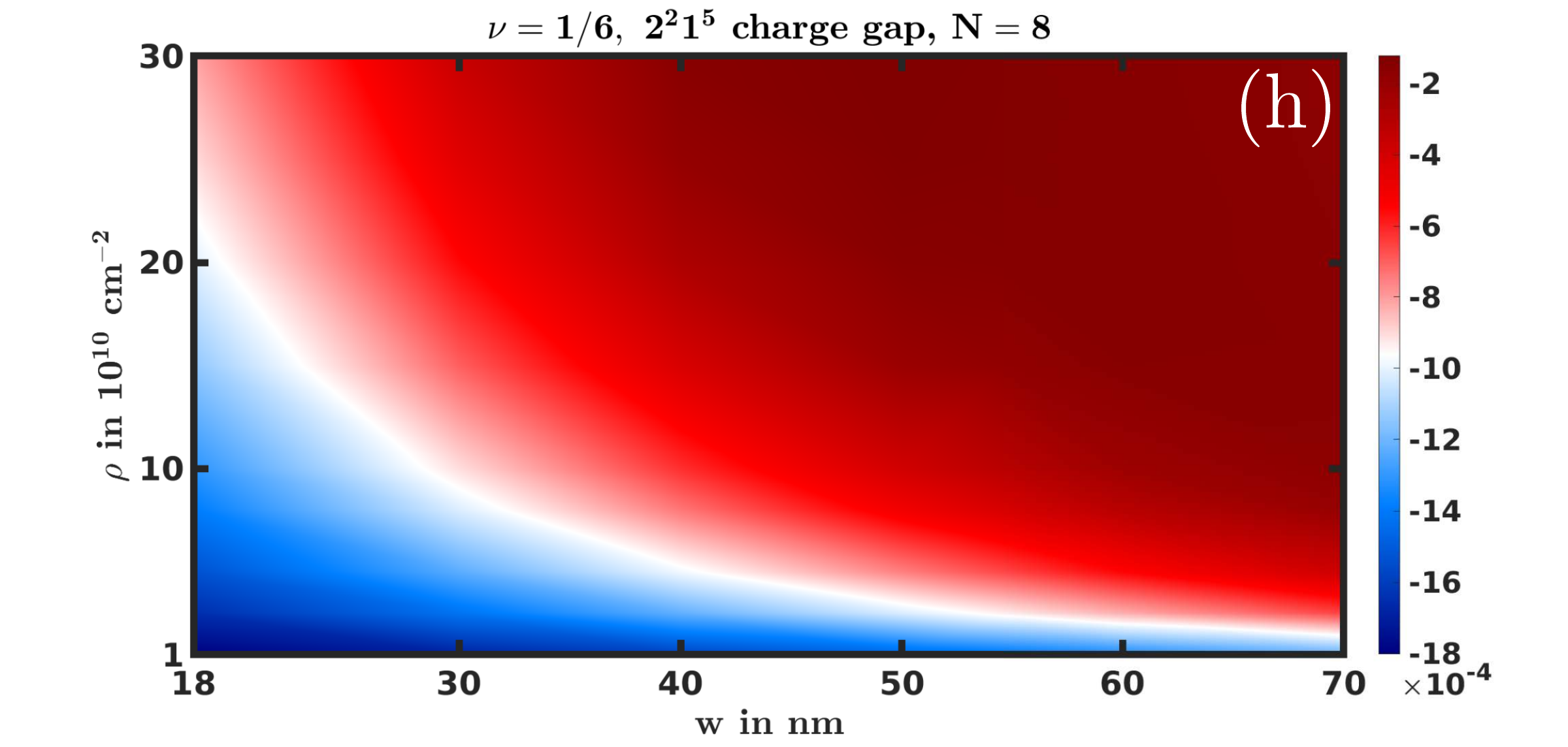}
			\includegraphics[width=0.32\textwidth]{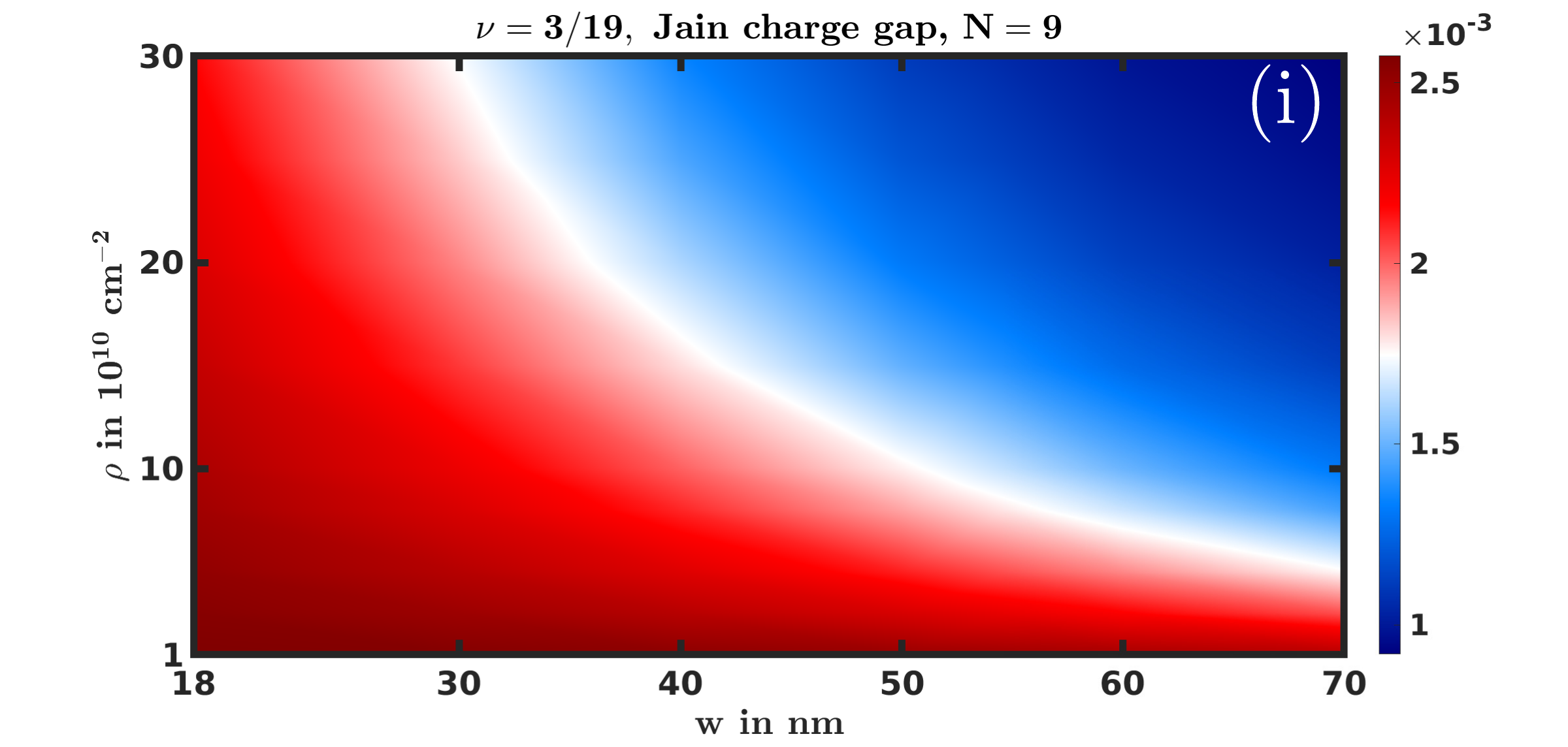}
		\end{center}
		\caption{
			Overlaps of the candidate states at $\nu{=}1/6$ with the exact lowest Landau level ground state [top panels (a), (b), and (c)], neutral gaps [middle panels (d), (e), and (f)] and charge gaps [bottom panels (g), (h), and (i)] evaluated in the spherical geometry for the $2^{2}1^{5}$ [left panels (a), (d), and (g)], $\bar{2}^{2}1^{7}$ [center panels (b), (e), and (h)], and CF Fermi sea [right panels (c), (f), and (i)] using the pseudopotentials of the finite-width interaction obtained using a local density approximation (LDA). All the left and center panels are for $N{=}8$ electrons. The $N{=}9$ filled-shell CFFS system at $\nu{=}1/6$ aliases with the Jain states at $\nu{=}3/19$ and $\nu{=}3/17$, and the charge gaps are calculated assuming the system forms an incompressible state at these Jain fractions.}
		\label{fig: 1_6_overlaps_gaps_finite_width_LDA}
	\end{figure*}

    The sine wave function is a simplified model for the transverse wave function in a QW, and, in particular, has no density dependence. Next, we model the effective interaction for electrons at various widths and densities of the QW using a LDA. In Figs.~\ref{fig: Laughlin_Jain_overlaps_gaps_finite_width_LDA} and \ref{fig: 1_6_overlaps_gaps_finite_width_LDA} we show the overlaps and gaps of the Laughlin, Jain, and candidate states at $\nu{=}1/6$ respectively, with the ED ground state obtained from LDA interaction. The overlaps of the $1/7$ Laughlin, $2/11$, $2/13$, and $3/19$ Jain, $2^{2}1^{5}$ and $\bar{2}^{2}1^{7}$ parton states with the ED ground states are upwards of $90{\%}$ for all the widths and densities considered. Generically, the $2^{2}1^{5}$ parton state has higher overlaps and gaps compared to the $\bar{2}^{2}1^{7}$ parton state. Here too, the ground state of the LDA interaction at the Moore-Read Pfaffian flux does not have $L{=}0$ for $N{=}8$. There are strong finite-size effects and aliasing issues, as evidenced by the gaps, that preclude us from arriving at a definitive conclusion with the ED results. 

    Only for the Laughlin state, the neutral gap is below the charge one, as it should be. The reason for this is that the finite-size effects for the Laughlin state are the weakest. The characteristic length scale, the effective magnetic length of CFs in the $ n/(2pn{\pm }1)$ Jain states, $\ell^{*}{=}\sqrt{2pn{\pm}1}\ell$, is the smallest for the Laughlin fractions in a given Jain sequence. Furthermore, the addition or removal of a single flux quantum only adds a single quasihole or quasiparticle in the Laughlin states, while for the $n/(2pn{\pm }1)$ Jain states, the addition or removal of a single flux quantum adds $n$ single quasiholes or quasiparticles. In small systems, especially for large $p$, these $n$ quasiparticles and quasiholes overlap and are therefore not sufficiently far apart from each other, inconsistent with the assumption made in the charge gap computation that these quasiparticles and quasiholes are non-interacting objects. We believe that only for the Laughlin state, the exact diagonalization computations reliably capture the charge gaps.  
	
		\section{Variational Monte Carlo calculations}
        \label{sec: VMC}
\begin{figure}[h]
		\begin{center}
			
			\includegraphics[width=0.43\textwidth,height=0.24\textwidth]{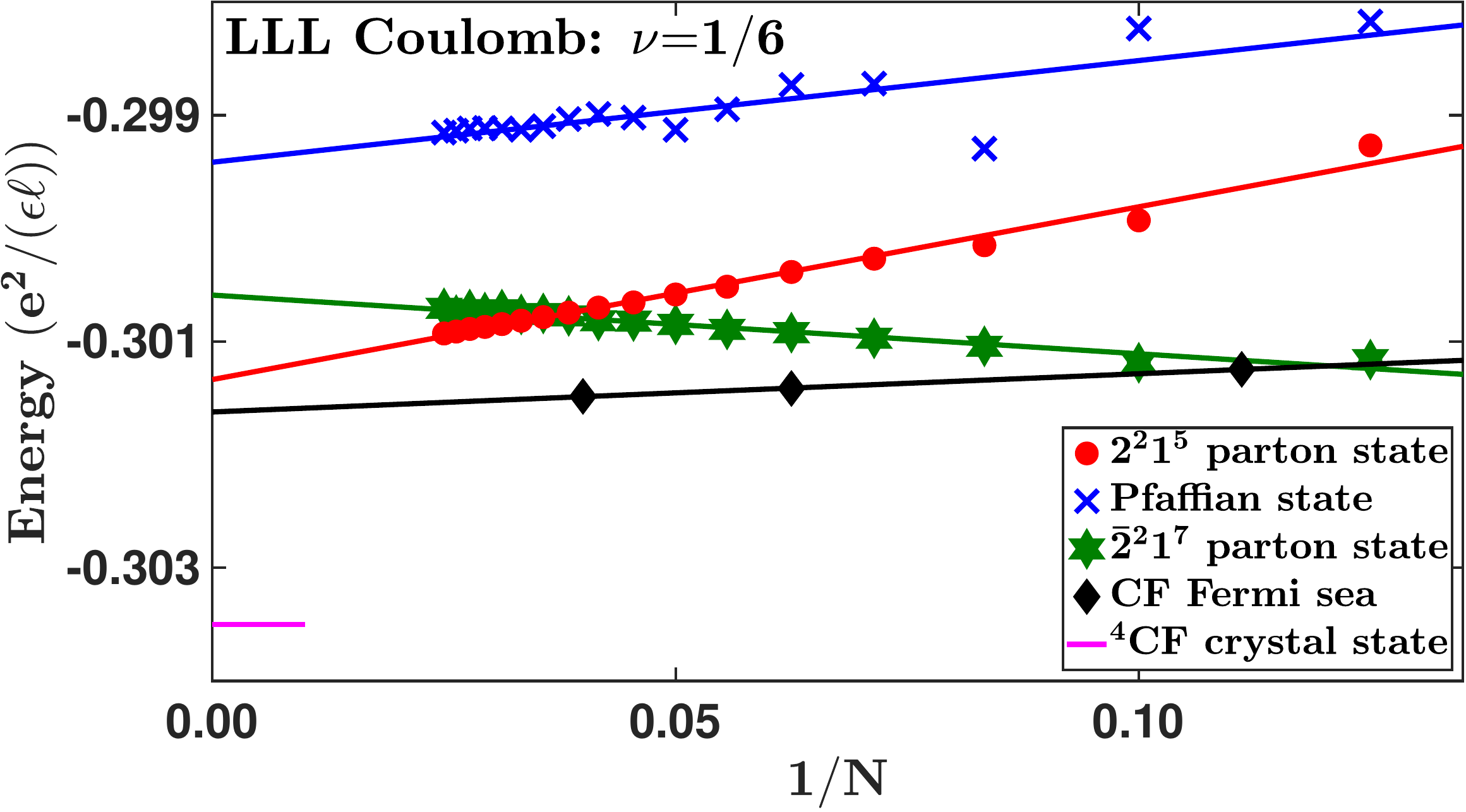} 
			\caption{(color online) Thermodynamic extrapolation of the per-particle Coulomb energies for the Pfaffian (blue crosses), $\bar{2}^{2}1^{7}$ (green heptagrams), $2^{2}1^{5}$ (red dots), the CF Fermi sea (CFFS, black diamonds) states at $\nu{=}1/6$ in the lowest Landau level of GaAs. The energies include the contribution of the positively charged background and are density-corrected~\cite{Morf86}. All energies are quoted in units of $e^2/(\epsilon\ell)$. The lines are linear fits in $1/N$. The thermodynamic energy of the $^{4}$CF crystal state (magenta line) is also shown for comparison.}
			\label{fig: extrapolations_energies_1_6}
		\end{center}
	\end{figure}

		In this section, we calculate the Coulomb energies of the candidate states outlined in Sec.~\ref{sec: wf_candidates}. For our variational calculations, we employ the JK~\cite{Jain97, Jain97b, Jain07, Moller05, Davenport12, Balram15a} projection scheme, which allows evaluation of energies for hundreds of electrons, leading to accurate thermodynamic limits. Fig.~\ref{fig: extrapolations_energies_1_6} shows thermodynamic limits for the energies of the CF Fermi sea (CFFS) and various paired CF states for an idealized system with zero thickness. Among the paired CF states, the $f$-wave state has the lowest energy, but it is slightly higher than the energy of the CFFS. The energy of the $^{4}$CFC at $\nu{=}1/6$ is ${-}0.3035$~\cite{Archer13}, which is lower than that of the CFFS, which has an energy of ${-}0.30162(1)$. This suggests that for the ideal zero-width system, the $^{4}$CFC may prevail. In contrast, previous calculations have shown that for the Coulomb interaction, the $1/7$ Laughlin and the $2/13$ Jain states provide excellent representations of the exact ground states~\cite{Zuo20}.

	%%%%%%%%%%%%%%%%%%%%%%%%%%%%%%%%%%%%%%%%%%%%%%%%%%%%%%%%%%%%%%%%%%%%%%%%%%%%%%
	%%%%%%%%%%%%%%%%%%%%%%%%%%%%%%%%%%%%%%%%%%%%%%%%%%%%%%%%%%%%%%%%%%%%%%%%%%%%%%
	
\begin{figure}[htpb]
	\begin{center}
	\includegraphics[width=0.45\textwidth]{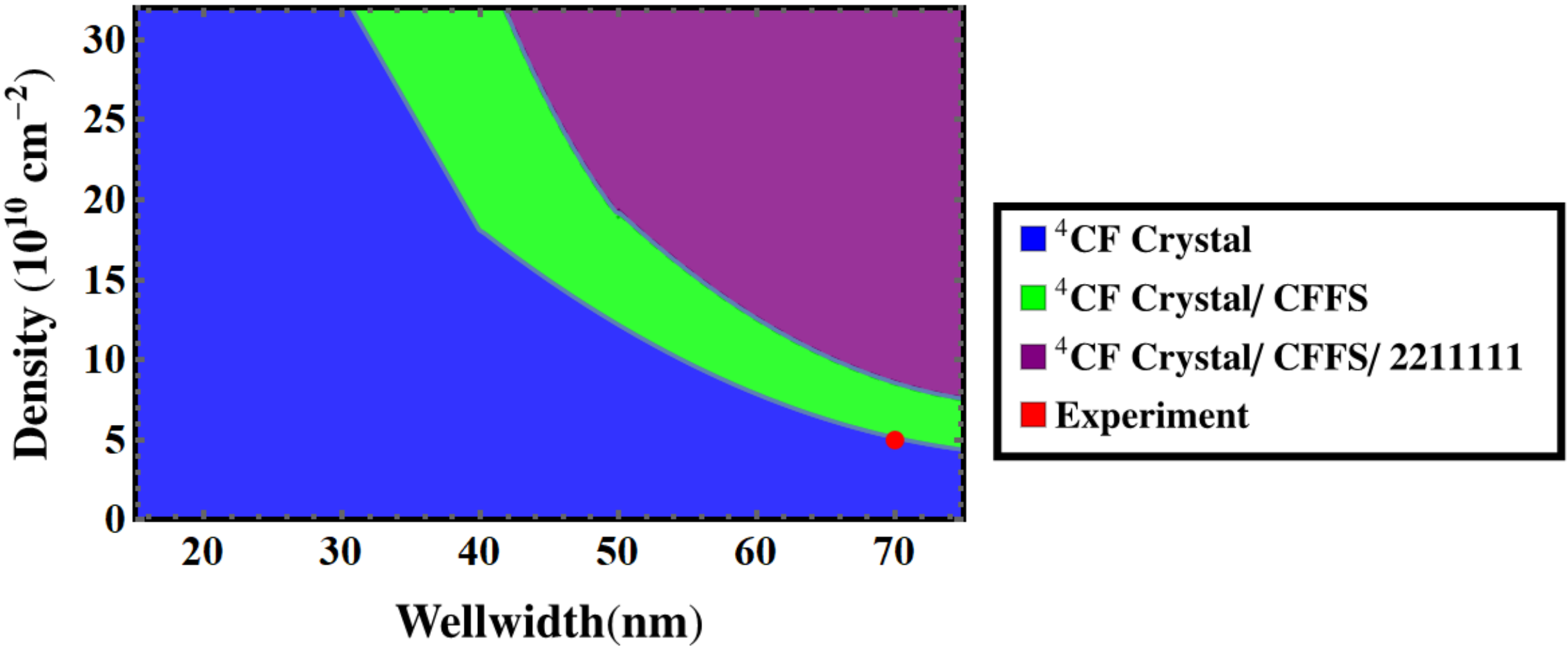}
    \caption{\label{fig: phase_diagram} Phase diagram depicting the competition between the four-vortex attached composite fermion crystal ($^{4}$CFC), the composite fermion Fermi sea (CFFS), and the $2211111$ state that represents $f$-wave pairing of composite fermions. A red square marks the parameters of the experiment~\cite{Wang25}.}
	\end{center}
	\end{figure}

	To obtain the phase diagram, we calculate the energies of the candidate states in the thermodynamic limit as a function of well width and density using the LDA interaction. For extrapolation to the thermodynamic limit, in addition to the density correction~\cite{Morf86}, we also take into account the corrections to the electron-background and background-background interactions due to finite thickness [see App.~\ref{app: background_subtraction_density_correction}]. The resulting phase diagram is shown in Fig.~\ref{fig: phase_diagram}. For low densities and narrow well widths (the blue region), the $^{4}$CFC has the lowest energy among the candidate states. For large densities and large well-widths (the purple region), the thermodynamic energies of 221111 parton state, the CFFS, and $^{4}$CFC are all within two standard deviations of each other, implying that their theoretical energies are too close to call. In the green region, the CFFS and the $^4$CFC are competitive. All other paired CF states have significantly higher energies in the thermodynamic limit. The parameters of the experiment of Ref.~\cite{Wang25} are shown by a red dot; for parameters in the vicinity of this point, all three states are competitive. Given that the various wave functions we are considering are approximate representations of the best states in the corresponding topological phases, the calculated phase boundaries are to be taken as approximate, and thus our study does not rule out a paired CF phase for the experimental parameters. 

	\section{Discussion}
    \label{sec: discussion}

A previous work~\cite{Zuo20} had suggested that for a zero-width sample, many of the Jain states are stabilized, but they are separated by CF crystals (CFCs). For weak disorder, the CFCs would be pinned, leading to insulating behavior, thus leading to FQHE states at $\nu{=}n/(6n{\pm}1)$ with insulators in between. For a higher disorder, when the FQHE states at $\nu{=}n/(6n{\pm}1)$ do not percolate, we have an insulator at all fillings, but the resistance has minima at $\nu{=}n/(6n{\pm}1)$ because of the presence of lakes of the FQHE liquid in the background of the CFC. In this work, we suggest that in wide QWs, FQHE also occurs at $\nu{=}1/6$, resulting in a phase diagram shown schematically in Fig.~\ref{fig: schematic_crystal_liquid_disorder}. 

Furthermore, our calculations show that the most plausible FQHE state at $\nu{=}1/6$ is the $2^2 1^5$ parton state, which represents an $f$-wave pairing of composite fermions~\cite{Balram19, Faugno19}. This has numerous experimentally verifiable consequences. As mentioned in the introduction, its thermal Hall conductance is predicted to be $\kappa_{xy}{=}c_{-}\pi^2k_B^2T/3h$ with $c_{-}{=}5/2$~\cite{Wen91, Balram19, Faugno19}. The $2^2 1^5$ parton state is expected to support two magnetoroton-like modes~\cite{Bose25}, one in which the particle-hole pair excitation resides in the factor of $\Phi_2$ and the other in which the excitation lives in the factor of $\Phi_1$~\cite{Balram21d, Balram24}. In the long-wavelength limit, these modes are referred to as gravitons. It is expected that for the $2^2 1^5$ state, these two gravitons have the same chirality~\cite{Liou19} as that of the graviton of the 1/3 Laughlin state~\cite{Bose25}. Recently, the graviton chirality has been measured for the 1/3 state using circularly polarized inelastic light-scattering~\cite{Liang24}.
    
	It is natural to seek generalizations of the $2^{2}1^{5}$ state. It may be viewed as a member of the sequence $n21^{5}$ described by the wave function
	\begin{equation}
	\Psi^{n21^{5}}_{2n/(11n+2)}=\mathcal{P}_{\rm LLL}\Phi_{n}\Phi_{2}\Phi^{5}_{1} \sim \Psi^{\rm Jain}_{n/(2n+1)}\Psi^{\rm Jain}_{2/5}\Phi_{1},
	\label{eq: parton_generalized_2_to_n}
	\end{equation}
	which produces incompressible states at $\nu{=}2n/(11n{+}2)$ at shift $\mathcal{S}^{n21^{5}}{=}n{+}7$. The first two members of the sequences, namely $n{=}1,2$ occur at $\nu{=}2/13$ and $1/6$ respectively, and have been observed in experiments~\cite{Pan02, Wang25}. A prediction would be that the next plateau would be observed at $\nu{=}6/35$ which is the $n{=}3$ member of the sequence given in Eq.~\eqref{eq: parton_generalized_2_to_n}. Note that the $6/35$ fraction is also the $n{=}6$ member of the tertiary Jain sequence $n/(6n{-}1)$, however, the $321^{5}$ state is topologically distinct from it.
	
	Another possibility is that the $2^{2}1^{5}$ state is the $n{=}1$ member of the parton sequence $n2^{2}1^{4}$ described by the wave function
	\begin{equation}
	\Psi^{n2^{2}1^{4}}_{n/(5n+1)}=\mathcal{P}_{\rm LLL}\Phi_{n}\Phi^{2}_{2}\Phi^{4}_{1} \sim \frac{\Psi^{\rm Jain}_{n/(2n+1)}[\Psi^{\rm Jain}_{2/5}]^{2}}{\Phi_{1}^{2}},
	\label{eq: parton_generalized_1_to_n}
	\end{equation}
	which produces incompressible states at $\nu{=}n/(5n{+}1)$ at shift $\mathcal{S}^{n21^{5}}{=}n{+}8$. The first two members of the sequences, namely $n{=}1,2$ occur at $\nu{=}1/6$ and $2/11$ respectively, and FQHE at these fractions has been observed in experiments~\cite{Pan02, Wang25}. However, we expect that the $2/11$ state in the LLL is the standard Jain CF type, which is topologically distinct from the $n{=}2$ member of the sequence given in Eq.~\eqref{eq: parton_generalized_1_to_n}. Therefore, we expect that the parton sequence of Eq.~\eqref{eq: parton_generalized_1_to_n} is unlikely to be stabilized in the LLL.
	
	Given that the second LL (SLL) and LLL primarily differ in the short-range part of the interaction, it is possible that the one-sixth (and also the quarter-filled~\cite{Faugno19}) state could be stabilized in the SLL. States lying in the Jain sequence described by composite fermions carrying a vorticity of $2p{>}2$ are topologically equivalent in the LLL and SLL~\cite{Balram21}. Interestingly, the half-filled state in the LLL \emph{at large widths} and the SLL is likely a $p$-wave paired state of composite fermions~\cite{Morf98, Rezayi17, Sharma21} while the FQHE states at $\nu{=}1/4$~\cite{Faugno19, Sharma23} and $\nu{=}1/6$ in wide quantum wells are predicted to be $f$-wave paired state of CFs.

	\begin{acknowledgments} 
		We thank Changyu Wang and Mansour Shayegan for numerous insightful discussions. A. C. B. thanks the Science and Engineering Research Board (SERB) of the Department of Science and Technology (DST) for financial support through the Mathematical Research Impact Centric Support (MATRICS) Grant No. MTR/2023/000002.  JKJ was supported in part by the U. S. Department of Energy, Office of Basic Energy Sciences, under Grant No. DE-SC-0005042. Computational portions of this research work were conducted using the Nandadevi supercomputer, which is maintained and supported by the Institute of Mathematical Science's High-Performance Computing Center, and the Advanced CyberInfrastructure computational resources provided by The Institute for CyberScience at The Pennsylvania State University. Some of the numerical calculations were performed using the DiagHam~\cite{diagham} package, for which we are grateful to its authors. We also extend our gratitude to the authors of the AQUILA package~\cite{Martin20}, which was used to obtain the LDA density profiles.
		
\end{acknowledgments}
	
\begin{appendix}

\section{Contribution of the positively charged background}
\label{app: background_subtraction_density_correction}
In spherical geometry, the background-background and electron-background contributions for zero thickness can be written as \cite{Jain07}:
\begin{equation}
E_{bb} + E_{eb} = -\frac{N^2}{2\sqrt{Q}} \frac{e^2}{\epsilon \ell}.
\end{equation}
This contribution is added to the electron-electron contribution to obtain the total energies of the states. We also include the effects of finite thickness in the electron-background and background-background interaction in our calculations. These can be computed using the wave functions in the transverse direction $\zeta(w)$ obtained from LDA~\cite{Sharma23}. To mitigate the $N$-dependence of the energies, we density-correct~\cite{Morf86} the finite-size results, i.e., multiply the per-particle energies by $\sqrt{2Q\nu/N}$, before extrapolating them to the thermodynamic limit. 

\end{appendix}

\bibliography{biblio_fqhe}
\end{document}